\documentclass[prl, 10pt, letterpaper, twocolumn, aps]{revtex4-2}
\usepackage{graphicx,amsmath,amssymb}
\usepackage{color}
\usepackage[normalem]{ulem}
\usepackage{epstopdf}
\usepackage{lmodern}
\usepackage[utf8]{inputenc}
\usepackage{lipsum}
\usepackage[colorlinks=true,citecolor=blue,linkcolor=blue,urlcolor=blue]{hyperref}
\usepackage{siunitx}
\DeclareSIUnit\gauss{G}

\definecolor{darkblue}{rgb}{0.0, 0.0, 0.55}

\def\ket#1{\left|#1\right\rangle}

\def\braket#1{\left\langle#1\right\rangle}
\def\ev#1{\left\langle#1\right\rangle}

\begin{document}
\title{Momentum entanglement for atom interferometry}

\author{F. Anders$^1$,  A. Idel$^1$, P. Feldmann$^2$, D. Bondarenko$^2$, S. Loriani$^1$, K. Lange$^1$, J. Peise$^1$, M. Gersemann$^1$, B.~Meyer$^1$, S. Abend$^1$, N. Gaaloul$^1$, C. Schubert$^{1,3}$, D. Schlippert$^1$, L. Santos$^2$, E. Rasel$^1$, and C. Klempt$^{1,3}$}
\affiliation{
$^1$ Institut f\"ur Quantenoptik, Leibniz Universit\"at Hannover,  Welfengarten~1, D-30167~Hannover, Germany \\
$^2$ Institut f\"ur Theoretische Physik,  Leibniz Universit\"at Hannover,   Appelstr. 2,  D-30167  Hannover,  Germany\\
$^3$\mbox{ Deutsches Zentrum für Luft- und Raumfahrt e.V.~(DLR),~Institut~für~Satellitengeod\"asie~und~Inertialsensorik,} c/o Leibniz Universit\"at Hannover,~DLR-SI, Callinstraße 36, 30167 Hannover, Germany
}

\begin{abstract}
Compared to light interferometers, the flux in cold-atom interferometers is low and the associated shot noise large.
Sensitivities beyond these limitations require the preparation of entangled atoms in different momentum modes.
Here, we demonstrate a source of entangled atoms that is compatible with state-of-the-art interferometers.
Entanglement is transferred from the spin degree of freedom of a Bose-Einstein condensate to well-separated momentum modes, witnessed by a squeezing parameter of \SI{-3.1(8)}{\dB}.
Entanglement-enhanced atom interferometers open up unprecedented sensitivities for quantum gradiometers or gravitational wave detectors.
\end{abstract}

\maketitle
Atom-interferometric measurements are fundamentally restricted by the Standard Quantum Limit (SQL), which can only be overcome by employing entangled atomic ensembles.
Surpassing the SQL with measurements based on internal degrees of freedom has been demonstrated in many different systems~\cite{Pezze2018} at room temperature~\cite{Wasilewski2010}, in ultracold ensembles ~\cite{Leroux2010,Louchet-Chauvet2010,Haas2014,Hosten2016}
and in Bose-Einstein condensates (BECs)~\cite{Gross2010,Riedel2010,Lucke2011,Ockeloen2013,Strobel2014,Muessel2014, Kruse2016, Zou2018}.
However, momentum-entangled sources as required for atom interferometers present a long-standing challenge.

\begin{figure}[b!]
\centering
\includegraphics[width=\columnwidth]{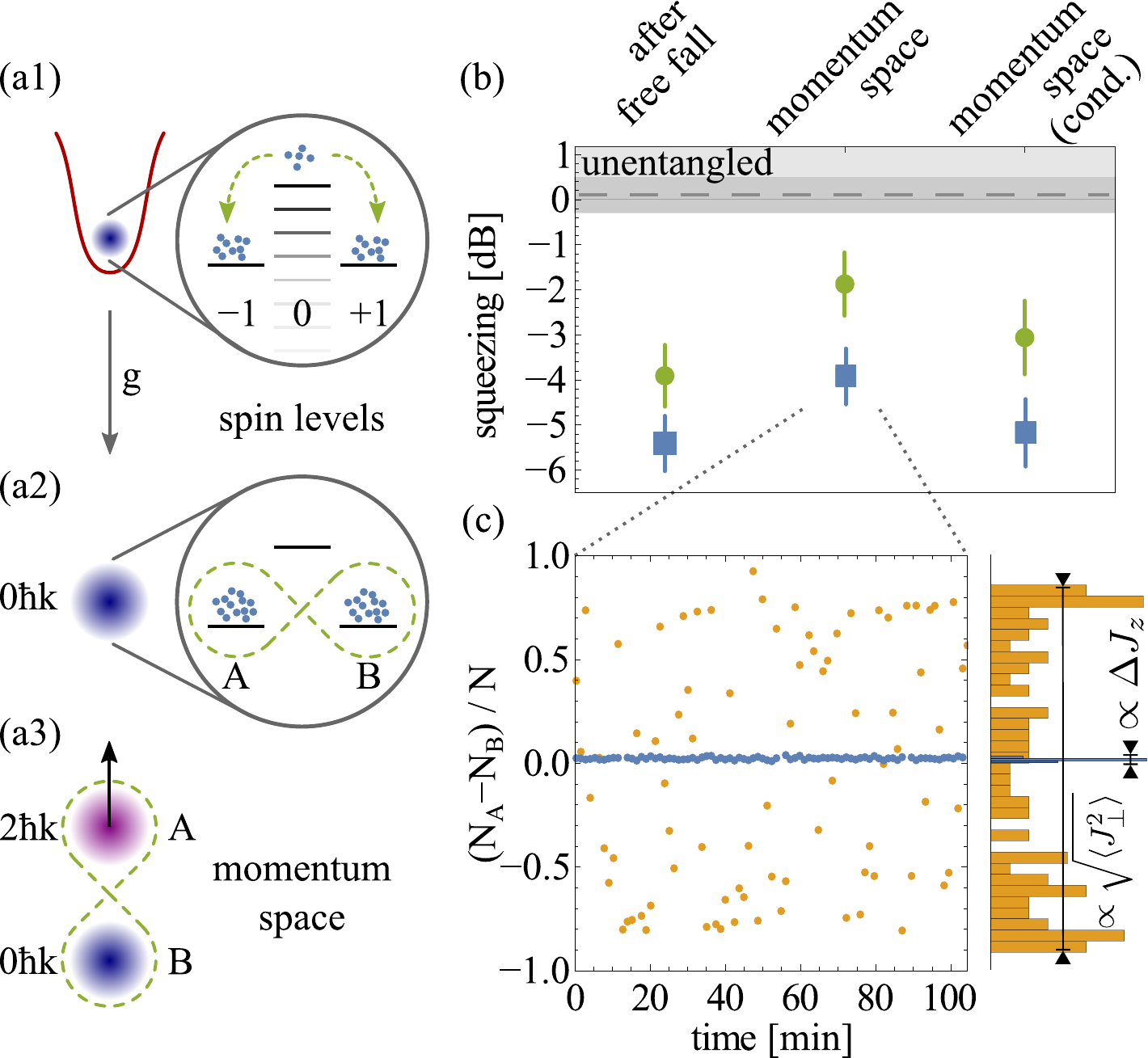}
\caption{
Entanglement in momentum space.
(a1) In a trap, spin-changing collisions create entangled atomic ensembles in two spin levels.
(a2) The entanglement is maintained during free fall and (a3) transferred to two distinct momentum modes. 
(b) Measured number squeezing $4 (\Delta J_z)^2/N$ (blue squares) and squeezing parameter~\cite{Hyllus2012, Lucke2014} (green dots) after free fall, in momentum space and conditionally (see text).
Values are well below the classical limit of \SI{0}{\dB}, which is experimentally verified with a coherent spin state (gray dashed line, uncertainty as dark gray area). 
(c) Measured atom number differences before (blue) and  after (orange) $\pi/2$-coupling.
Before the coupling, the two modes A and B are equally populated and yield ultra-low fluctuations in the number difference.
After the coupling, the fluctuations are large, with a characteristic cumulation at extreme values. Each set of data points in (b) is derived from such data.
}
\label{fig1}
\end{figure}

\begin{figure*}[ht!]
\centering
\includegraphics[width=1.9\columnwidth]{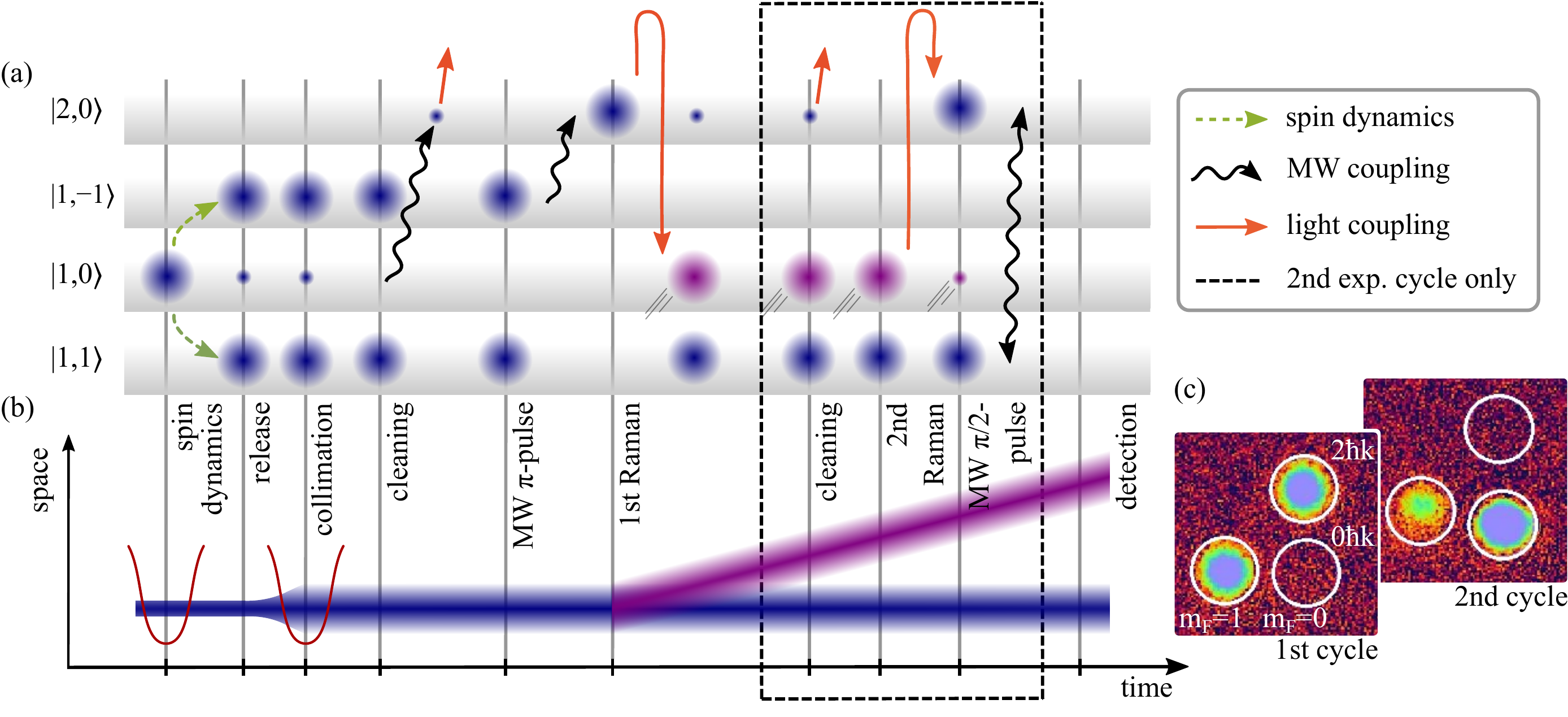}
\caption{Schematic overview of the experimental sequence.
(a) The transfer of atoms between the various involved spin states.
The type of coupling is indicated by different arrows (legend). We measure $J_z$ and $J_\perp$ in two alternating measurement cycles.
Operations enclosed by the dashed rectangle only take place in the second experimental cycle to measure $J_\perp$.
(b) The space-time diagram shows the effect on the spatial mode of the atomic cloud in the free-falling reference frame (only depicting the first experimental cycle, time axis not to scale).
(c) Typical absorption images taken at the end of both cycles.}
\label{fig2}
\end{figure*}

Controlled atomic collisions were shown to enable the generation of entanglement between spatial modes~\cite{Esteve2008,Berrada2013,Lange2018,Fadel2018,Kunkel2018}, as well as correlated and entangled atomic pairs in momentum space~\cite{Bucker2011, Kheruntsyan2012, Shin2019}.
Further schemes for the generation of entanglement between momentum modes have been proposed theoretically~\cite{Salvi2018,
Geiger2018,
Shankar2019,
Szigeti2020}.
However, both the momentum and the spatial mode of the atoms are often determined by the generation process, such that an integration into state-of-the-art atom interferometers would be a challenging task.
An alternative approach, the transfer of entanglement from internal to external degrees of freedom, was so far only proposed theoretically~\cite{Szigeti2020a}.

In this Letter, we report on a source for momentum-entangled atoms featuring the excellent mode quality of a Bose-Einstein condensate (BEC).
We achieve this by
the transfer of highly entangled twin-Fock states in the spin degree of freedom of a BEC to momentum space (Fig.~\ref{fig1}).
The twin-Fock states are created in a trap and released into free space, where one of the twin modes is coherently transferred to a well-separated momentum mode.
Between the two momentum modes, we record number and phase fluctuations and obtain a spin squeezing parameter~\cite{Hyllus2012, Lucke2014} of \SI{-3.1(8)}{\dB}, which proves entanglement in momentum space.
The demonstrated entanglement concept is directly applicable in existing atom interferometers to enable sensitivities beyond the SQL.
Such a quantum-enhanced resolution is of vital interest for future large-scale atom interferometers which measure relative observables, for Earth observation gradiometry~\cite{Snadden1998}, for tests of the Einstein Equivalence Principle~\cite{Aguilera2014a} and for the proposed gravitational wave detectors~\cite{Tino2019,Canuel2018, Canuel2020}.
Momentum-entangled atoms further constitute a promising probe for tests of fundamental decoherence~\cite{Schrinski2020} and tests of Bell nonlocality with massive particles~\cite{Laloe2009}.

We initiate our experiments by the preparation of high-quality entangled states in spin space. 
A BEC of $10^4$ $^{87}$Rb atoms is produced in a crossed-beam optical dipole trap with
trapping frequencies of $2\pi \times(150,160, 220)\,$Hz.
The atoms are prepared in the hyperfine level $\ket{F,m_F}=\ket{1,0}$ at an actively stabilized, homogeneous magnetic field of \SI{0.73}{\gauss} oriented parallel to the gravitational acceleration.
We employ spin-changing collisions~\cite{Luecke2011a, Gross2011, Hamley2012} to generate highly entangled twin-Fock states $\ket{N_A=N/2}\otimes \ket{N_B=N/2}$ in the two levels $m_F=\pm1$.
Following earlier work~\cite{Zhang2013,Luo2017,Feldmann2018}, we generate these states by a quasi-adiabatic crossing of a quantum phase transition.
In our realization, we apply an intensity-stabilized homogeneous microwave (MW) field which is blue detuned by \SI{400}{\kilo \Hz} from the transition $\ket{1,0}\leftrightarrow\ket{2,0}$ and linearly ramp the field intensity.
Without MW dressing, an atom pair in $\ket{1,\pm1}$ has a relative energy of $q=h\times$\SI{38.5}{\Hz}$/\text{atom}$ compared to a pair in $\ket{1,0}$ due to the quadratic Zeeman shift.
For the initial spin orientation, the BEC in $\ket{1,0}$ is thus in the many-body ground-state of the system.
We then apply a \SI{1020}{\ms} linear intensity ramp to the dressing field, which lowers the energy of the $\ket{1,\pm 1}$ levels to $-h\times$\SI{5}{\Hz} each~\footnote{See Supplemental Material at [URL will be inserted by the publisher] for details on the quasi-adiabatic state preparation, including references~\cite{Zhang2013, Luo2017, Feldmann2018, Chang2004, Leslie2009, Klempt2010, Linnemann2016a}.}.
The atoms follow the ground-state of the system towards a twin-Fock state at the end of the ramp.
Despite experimental noise and finite ramping speed, \SI{93 \pm 5}{\percent} of the atoms are transferred to the levels $\ket{1,\pm1}$.
The overall preparation yields a total of $\ev{N}=9300$ atoms with only \SI{10}{\percent} relative fluctuations, which are prepared in a highly entangled twin-Fock state in the spin degree of freedom.

\begin{figure}[t!]
\centering
\includegraphics[width=\columnwidth]{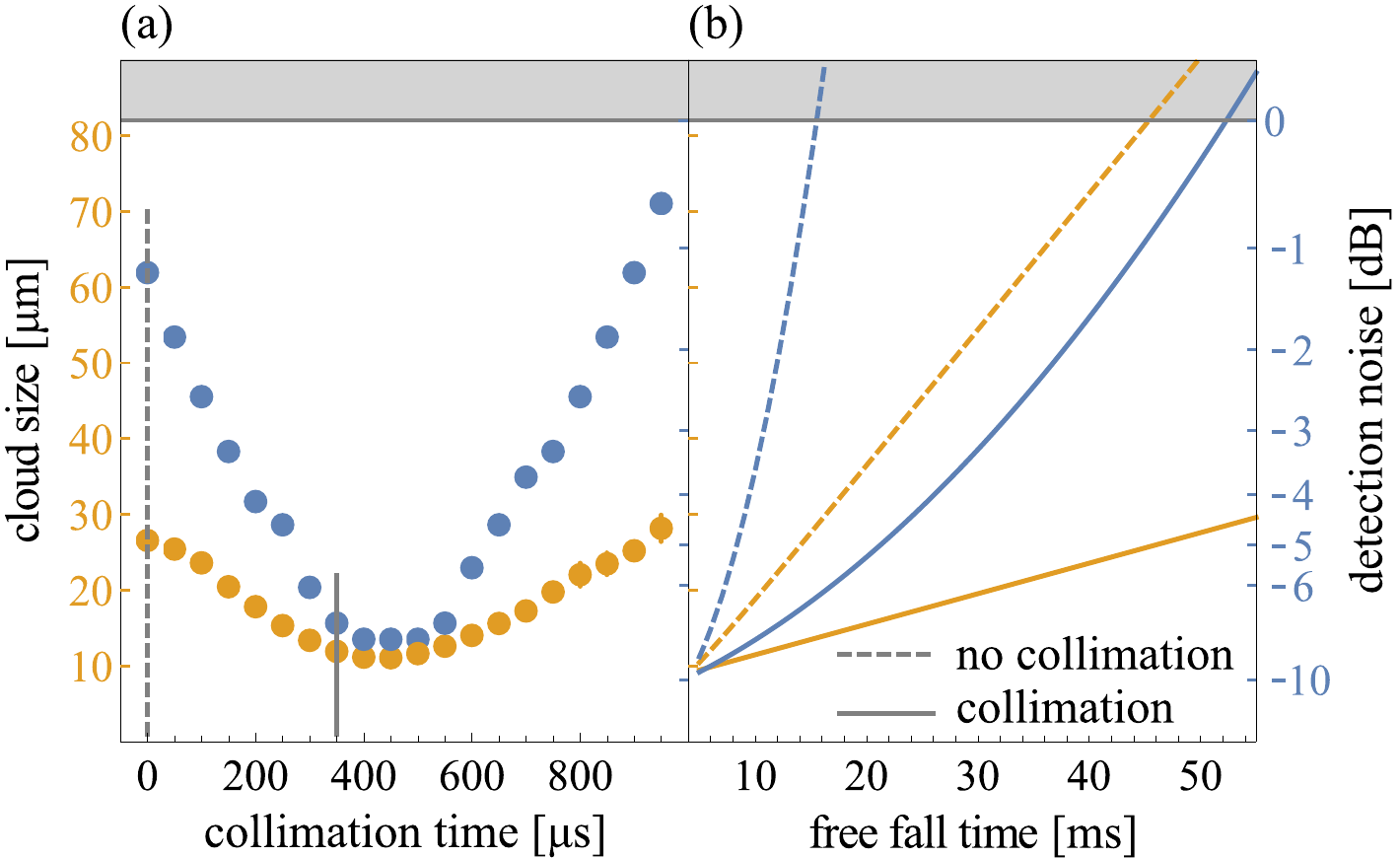}
\caption{
Effect of the collimation on cloud size (orange, left scale) and detection noise (blue, right scale).
(a) At a fixed free-fall time of \SI{13}{\ms} after the collimation, the size of the atomic cloud is measured as a function of the collimation pulse time.
The cloud size determines the minimal number of relevant pixels on the CCD camera. This corresponds to a minimal detection noise, which we compare to shot-noise (gray area).
(b) Extrapolation to longer free-fall times based on expansion rates measured for the two settings marked by vertical lines in (a).
The maximal free-fall time allowing for sub-shot-noise detection is increased by a factor of three to about $50$~ms.
}
\label{fig4}
\end{figure}

The transfer to momentum space requires a release to free space without destruction of the entanglement (the full sequence is displayed in Fig.~\ref{fig2}).
The trapping laser fields are switched off instantaneously to initiate a free expansion which is dominated by mean-field interaction~\cite{Castin1996}.
This accelerated expansion turns quickly into a ballistic expansion after the density has dropped.
Because of the initial high density, necessary to generate entanglement via spin-changing collisions, the outcoupled cloud has a broad velocity distribution of \SI{1.8}{\milli \m \per \s}.
However, a narrow velocity distribution is favorable for the acceleration by a stimulated Raman transition to avoid Doppler shifts (along the Raman-beam direction) and the sensing of phase and intensity gradients (along transverse directions).
Furthermore, a broad velocity distribution would be converted into an extended spatial distribution.
We therefore apply a three-dimensional collimation pulse~\cite{Ammann1997}.
After \SI{1}{\ms} of free fall, we flash the dipole trapping field with its original strength for an adjustable duration.
Figure~\ref{fig4} shows the effect of this collimation.
For an increasing collimation pulse length, the detected cloud size is first reduced, reaches a minimum, and increases again.
In our experiments, we choose a pulse length of \SI{350}{\micro\second} to avoid refocusing and the corresponding inflection of the cloud.
The reduction of the cloud size is also essential for its detection with sub-shot-noise sensitivity~\cite{Lucke2014}.
Extended clouds require more pixels on the final absorption images and thereby sample more noise.
Without collimation, the detection noise would remain at suitably low values only for a few milliseconds of free-fall time.
In our experiments, the collimation reduces the detection noise from \SIrange{-0.2}{-6.2}{\dB} at our typical free-fall time of \SI{16}{\milli \s}, and therefore actually enables a transfer of entanglement to momentum space and its subsequent detection.

After the collimation, the clouds slowly expand for another \SI{2.5}{\milli \s} to be sufficiently dilute to remove the remaining atoms from the level $\ket{1,0}$ by a MW transfer and a resonant light pulse.
We detect no leftover atoms and, after another MW transfer (Fig.~\ref{fig2}), a clean, free-falling twin-Fock state in the levels $\ket{1,1}$ and $\ket{2,0}$ remains.

We evaluate the quality of the twin-Fock state of spin-levels after collimated free fall of \SI{15}{\ms}.
Analogous to prior work~\cite{Lucke2011,Lucke2014}, we detect the number of atoms $N_{\text{A}/\text{B}}$ in the two modes $\ket{0 \hbar k; 1,1}$ and $\ket{0 \hbar k; 2,0}$, and observe strongly reduced fluctuations.
Fig.~\ref{fig1}~(b) shows the obtained number squeezing $4 (\Delta J_z)^2/N$ of \SI{5.4(6)}{\dB} below shot noise.
A detection of entanglement requires the measurement of a conjugate observable such as the relative phase.
Here, the phase can be observed after performing a symmetric $\pi/2$ MW coupling pulse between the two modes.
The two measurements are combined in a squeezing parameter~\cite{Hyllus2012, Lucke2014} $\xi^2=\left( \Delta J_z \right)^2 / \left(2 \ev{J_\perp^2/(\hat{N}-1)}-\ev{(\hat{N}/2)/(\hat{N}-1)} \right)$, where $(\Delta J_z)^2$ represents the variance of the number difference $J_z=\frac{1}{2}(\hat{N}_A-\hat{N}_B)$ and $\braket{J_\perp^2}$ is the second moment of the same number difference after the $\pi/2$ coupling (Fig.~\ref{fig1}~(c)).
The squeezing parameter proves entanglement if $\xi^2<1$.
From our measurements in free fall, we obtain a squeezing parameter of \SI{-3.9(7)}{\dB} with respect to the classical bound.
The obtained squeezing in free fall is worse than in the trapped case~\cite{Lucke2014}, which results from an increased detection noise (enlarged cloud and technical noise), and does not imply a deterioration of the state.
The reduced fluctuations after rotation (\SI{69}{\percent} of the ideal twin-Fock value of $\ev{J_\perp^2}=N/2(N/2+1)$) can be explained by decoherence due to longer holding times in the trap and asymmetries of the collimation procedure, which may lead to non-identical spatial phase patterns for the two modes.
However, we obtain a clear signal of entanglement in free-falling BECs, which presents a central result of this publication.
In complementary work, squeezed samples of thermal atoms were successfully released to a free fall of \SI{8}{\milli \s}~\cite{Malia2020}.

\begin{figure}[t!]
\centering
\includegraphics[width=\columnwidth]{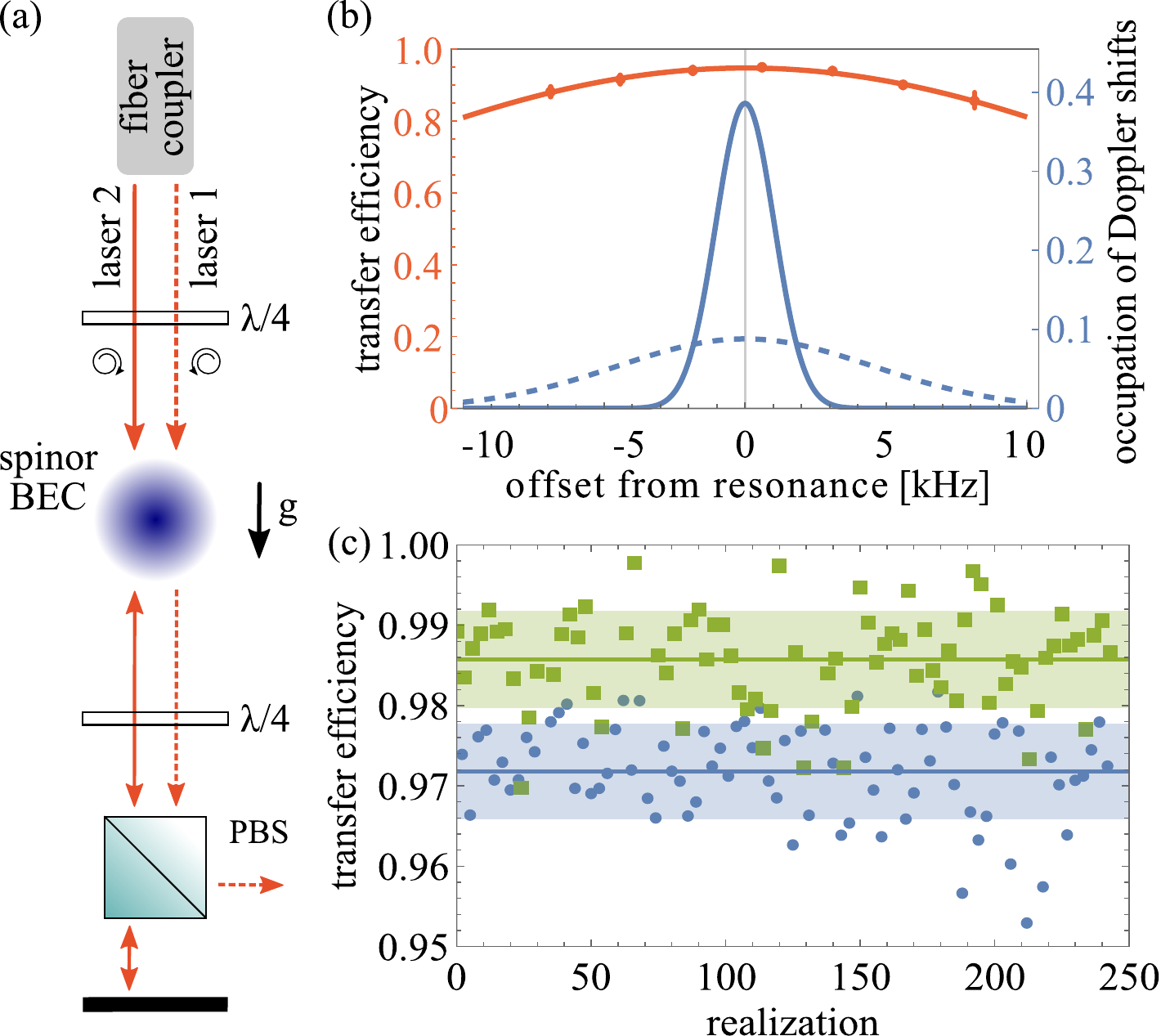}
\caption{
Design and characterization of the Raman coupling scheme.
(a) Schematic of the optical set-up for realizing Raman transitions. Two oppositely circular-polarized  phase-locked Raman beams (laser 1 and 2) pass the atomic cloud from above. Due to selection by a polarizing beam splitter (PBS), only laser 2 is retro-reflected. Thereby, only one pair of beams enables a momentum transfer of $2 \hbar k$ and unwanted transfers are suppressed. To allow for future measurements of gravity, the setup is aligned with the gravitational accelaration $g$.
(b) Raman spectroscopy of the clock transition. The experimental data of the spectroscopy (orange data points and fit) is compared to the distributions of Doppler shifts due to the velocity spread before and after collimation (blue dashed and solid line).
(c) Transfer efficiencies for two consecutive Raman pulses from $\ket{2,0}$ to $\ket{1,0}$ (blue circles) and back (green dots).
}
\label{fig3}
\end{figure}

The central development for achieving momentum entanglement is a high-efficiency momentum transfer.
This is achieved with resonant Raman laser pulses which couple the levels $\ket{2,0}$ and $\ket{1,0}$ by a two-photon transition with \SI{1.1}{\giga \Hz} red-detuning from the $5 P_{3/2}$ manifold.
The pulses are temporally shaped with $\sin^2$ edges to reduce the frequency sensitivity in Fourier space.
Two separate diode lasers are used where the phase of the laser that couples to $\ket{1,0}$ (laser 1) is stabilized to the $\ket{2,0}$ laser (laser 2)~\cite{Berg2015}.
The phase-stabilized beams are superposed with crossed linear polarizations, and mode-cleaned by an optical fiber.
After the first fiber, the beam is switched by a single acousto-optical modulator and delivered to the experimental chamber via a second optical fiber.
The intensity ratio is adjusted to a value of $ I_2/I_1=0.93$ (in front of the atoms), where the AC Stark shifts induced by both frequencies compensate, such that the Raman coupling is insensitive to fluctuations of the total power. 
After outcoupling along the vertical direction, the Raman beams are given opposite circular polarizations and pass the falling cloud (Fig.~\ref{fig3}~(a)).
Below the cloud, laser beam 1 is removed, and laser beam 2 is reflected back to the atoms.
The combination of laser 1 from above and laser 2 from below enables an upward acceleration by two photon recoil quanta (\SI{11.8}{\milli \m \per \s}) that is associated with a spin transfer from  $\ket{1,0}$ to $\ket{2,0}$.
The obtained change of velocity is much larger than the velocity distribution of the cloud with an rms width of \SI{0.4}{\milli \m \per \s}, enabling a clean preparation of distinct momentum modes.
The Raman pulses are applied after a free-fall time of \SI{7.7}{\milli \s}, because the gravitational acceleration to \SI{76}{\milli \m \per \s} provides a sufficient Doppler shift to suppress unwanted transitions due to imperfect polarization and reflection.

We validate the efficiency of the Raman coupling by applying it to a free-falling BEC in the level $\ket{2,0}$.
Figure~\ref{fig3}~(b) shows a spectroscopy of the Raman transition (orange) and compares it to the Doppler shifts due to the residual velocity spread (blue).
The collimation reduces the ballistic expansion by \SI{77}{\percent} corresponding to a Doppler spread of \SI{1}{\kilo \Hz} (less than \SI{2}{\percent} of the Fourier width of the Raman pulse), equivalent to an effective temperature of \SI{850}{\pico \K}.
The residual expansion rate is sufficiently small not to reduce the efficiency of the Raman coupling.
Figure~\ref{fig3} (c) shows the transfer efficiency for a transition from $\ket{0 \hbar k; 2,0}$ to $\ket{2 \hbar k; 1,0}$ (upward acceleration, blue) and a subsequent transition back to $\ket{0 \hbar k; 2,0}$ (downward acceleration, green).
The transfer pulses yield an efficiency of \SI{97.2 \pm 0.6}{\percent} and \SI{98.5 \pm 0.6}{\percent}, respectively.
We attribute the efficiency limitation to two main effects:
(i) Because of finite temperature, there will be a small fraction of atoms with larger velocities which are not transferred due to the Doppler shift.
Characteristically, this effect is strongly reduced for the second pulse, where the fast atoms have already been removed.
(ii) Relative drifts of the Raman beam intensities, as observed in our experiment, drive the system away from the ideal AC-Stark-suppression.
Therefore, depending on the elapsed time since the last calibration, the intensity fluctuations start to couple more to the resonance frequency, eventually reducing the efficiency.
This effect is relevant for many hours of measurements and could be circumvented by an improved intensity stabilization in the future.
However, the recorded efficiencies belong to the best reported Raman transfers~\cite{Butts2013, Kotru2015, Jaffe2018} and constitute the main technical achievement to successfully transfer entangled states to different momentum modes~\footnote{Other techniques, such as Bragg transitions or Bloch oscillations~\cite{Kovachy2012, Abend2016}, can reach even better efficiencies but cannot be applied within our scheme.}.
Note that we take all atoms of the prepared state into account, without any velocity selection before the momentum transfer.

The described concepts can now be combined to prove entanglement in momentum space. 
We apply the Raman transfer to our Twin Fock state by coupling atoms in $\ket{0 \hbar k; 2,0}$ to a finite momentum state $\ket{2 \hbar k; 1,0}$.
After an additional time of flight of \SI{7.6}{\ms}, we detect two clouds clearly separated by \SI{80 \pm 1}{\micro \m} (center of mass).
A strong magnetic field gradient in horizontal direction enables an independent detection of the unaffected atoms in $\ket{0 \hbar k; 1,1}$ and the small amount of leftover atoms in $\ket{0 \hbar k; 2,0}$ that stems from the imperfect Raman transfer.
For the two macroscopically occupied clouds that drift apart, we record \SI{-3.9(6)}{\dB} number squeezing.
If the measurement of the leftover atoms is exploited to predict the measurement outcome, thereby creating a conditional Dicke state, we obtain a number squeezing of \SI{-5.2(7)}{\dB}.
In order to record the phase difference as a conjugate observable, we reverse the momentum transfer before the clouds separate substantially.
Within \SI{40}{\micro \s} after the first Raman transfer, another cleaning procedure removes the leftover atoms in $\ket{0 \hbar k;2,0}$ and a second Raman coupling decelerates the atoms back to $\ket{0 \hbar k; 2,0}$.
Now, it is possible to couple the two twin-Fock modes by a MW $\pi/2$ pulse.
Again, we obtain extremely large fluctuations in the number difference, with a corresponding second moment of $\ev{J_\perp^2}=0.63(5) \times N/2(N/2-1)$, and calculate a squeezing parameter of \SI{-1.9(7)}{\dB}.
For the conditional case, we obtain a squeezing parameter of \SI{-3.1(8)}{\dB} (Fig.~\ref{fig1}~(b) middle and right set of data points).
This proof of entanglement between two atomic modes, well-separated in momentum space, presents our main result.

The observed entangled states are directly applicable for inertially sensitive atom interferometry beyond the SQL.
The desired quantum-enhanced phase sensitivity can be obtained by a $\pi/2$ coupling pulse between the two twin-Fock modes (representing the first beam splitter of the interferometer) before the Raman transfer.
It is a characteristic advantage of the presented approach that these coupling pulses can be performed in the well-controlled spin space.
The presented scheme is of course not limited to twin-Fock states, but also applies to other entangled states in spin space, for example spin-squeezed states~\cite{Hamley2012,Kruse2016}.
The demonstrated source of entangled, Bose-condensed atoms in momentum space opens the path to operate future atom interferometers with quantum-enhanced sensitivities.
This is specifically desirable for relative measurements with multiple atom interferometers, where some dominant technical noise sources like vibrational noise are suppressed by common-noise rejection.
Targeted interferometer applications include gradiometers for Earth observation~\cite{Snadden1998}, tests of the Einstein Equivalence Principle~\cite{Dimopoulos2007, Hartwig2015,  Asenbaum2020}
and future terrestric~\cite{Schubert2019, Canuel2020} and space-borne~\cite{Dimopoulos2008,Hogan2011a,Loriani2019, ElNeaj2020} gravitational wave detectors.
The extreme sensitivity of entangled states can be employed for testing collapse theories~\cite{Schrinski2020}.
Finally, the achievable spatial separation enables testing local realism with ultracold atomic ensembles~\cite{Laloe2009}.

We thank A. Smerzi and G. T\'o{}th for valuable discussions and J. Arlt for a critical review of the manuscript.
We   acknowledge   support   from  the European Union through the QuantERA grant 18-QUAN-0012-01 (CEBBEC).
The work is funded by the Deutsche Forschungsgemeinschaft (DFG) under Germany's Excellence Strategy (EXC-2123 QuantumFrontiers 390837967), and through CRC 1227 (DQ-mat), projects A02 and B07. F.A. acknowledges support from the Hannover School for Nanotechnology (HSN). D.S. acknowledges support by the Federal Ministry of Education and Research (BMBF) through the funding program Photonics Research Germany under contract number 13N14875.

\bibliography{main}

\begin{thebibliography}{65}%
\makeatletter
\providecommand \@ifxundefined [1]{%
 \@ifx{#1\undefined}
}%
\providecommand \@ifnum [1]{%
 \ifnum #1\expandafter \@firstoftwo
 \else \expandafter \@secondoftwo
 \fi
}%
\providecommand \@ifx [1]{%
 \ifx #1\expandafter \@firstoftwo
 \else \expandafter \@secondoftwo
 \fi
}%
\providecommand \natexlab [1]{#1}%
\providecommand \enquote  [1]{``#1''}%
\providecommand \bibnamefont  [1]{#1}%
\providecommand \bibfnamefont [1]{#1}%
\providecommand \citenamefont [1]{#1}%
\providecommand \href@noop [0]{\@secondoftwo}%
\providecommand \href [0]{\begingroup \@sanitize@url \@href}%
\providecommand \@href[1]{\@@startlink{#1}\@@href}%
\providecommand \@@href[1]{\endgroup#1\@@endlink}%
\providecommand \@sanitize@url [0]{\catcode `\\12\catcode `\$12\catcode
  `\&12\catcode `\#12\catcode `\^12\catcode `\_12\catcode `\%12\relax}%
\providecommand \@@startlink[1]{}%
\providecommand \@@endlink[0]{}%
\providecommand \url  [0]{\begingroup\@sanitize@url \@url }%
\providecommand \@url [1]{\endgroup\@href {#1}{\urlprefix }}%
\providecommand \urlprefix  [0]{URL }%
\providecommand \Eprint [0]{\href }%
\providecommand \doibase [0]{https://doi.org/}%
\providecommand \selectlanguage [0]{\@gobble}%
\providecommand \bibinfo  [0]{\@secondoftwo}%
\providecommand \bibfield  [0]{\@secondoftwo}%
\providecommand \translation [1]{[#1]}%
\providecommand \BibitemOpen [0]{}%
\providecommand \bibitemStop [0]{}%
\providecommand \bibitemNoStop [0]{.\EOS\space}%
\providecommand \EOS [0]{\spacefactor3000\relax}%
\providecommand \BibitemShut  [1]{\csname bibitem#1\endcsname}%
\let\auto@bib@innerbib\@empty
\bibitem [{\citenamefont {Pezz\`e}\ \emph {et~al.}(2018)\citenamefont
  {Pezz\`e}, \citenamefont {Smerzi}, \citenamefont {Oberthaler}, \citenamefont
  {Schmied},\ and\ \citenamefont {Treutlein}}]{Pezze2018}%
  \BibitemOpen
  \bibfield  {author} {\bibinfo {author} {\bibfnamefont {L.}~\bibnamefont
  {Pezz\`e}}, \bibinfo {author} {\bibfnamefont {A.}~\bibnamefont {Smerzi}},
  \bibinfo {author} {\bibfnamefont {M.~K.}\ \bibnamefont {Oberthaler}},
  \bibinfo {author} {\bibfnamefont {R.}~\bibnamefont {Schmied}},\ and\ \bibinfo
  {author} {\bibfnamefont {P.}~\bibnamefont {Treutlein}},\ }\bibfield  {title}
  {\bibinfo {title} {Quantum metrology with nonclassical states of atomic
  ensembles},\ }\href {https://doi.org/10.1103/RevModPhys.90.035005} {\bibfield
   {journal} {\bibinfo  {journal} {Rev. Mod. Phys.}\ }\textbf {\bibinfo
  {volume} {90}},\ \bibinfo {pages} {035005} (\bibinfo {year}
  {2018})}\BibitemShut {NoStop}%
\bibitem [{\citenamefont {Wasilewski}\ \emph {et~al.}(2010)\citenamefont
  {Wasilewski}, \citenamefont {Jensen}, \citenamefont {Krauter}, \citenamefont
  {Renema}, \citenamefont {Balabas},\ and\ \citenamefont
  {Polzik}}]{Wasilewski2010}%
  \BibitemOpen
  \bibfield  {author} {\bibinfo {author} {\bibfnamefont {W.}~\bibnamefont
  {Wasilewski}}, \bibinfo {author} {\bibfnamefont {K.}~\bibnamefont {Jensen}},
  \bibinfo {author} {\bibfnamefont {H.}~\bibnamefont {Krauter}}, \bibinfo
  {author} {\bibfnamefont {J.~J.}\ \bibnamefont {Renema}}, \bibinfo {author}
  {\bibfnamefont {M.~V.}\ \bibnamefont {Balabas}},\ and\ \bibinfo {author}
  {\bibfnamefont {E.~S.}\ \bibnamefont {Polzik}},\ }\bibfield  {title}
  {\bibinfo {title} {Quantum {n}oise {l}imited and {e}ntanglement-{a}ssisted
  {m}agnetometry},\ }\href {https://doi.org/10.1103/PhysRevLett.104.133601}
  {\bibfield  {journal} {\bibinfo  {journal} {Phys. Rev. Lett.}\ }\textbf
  {\bibinfo {volume} {104}},\ \bibinfo {pages} {133601} (\bibinfo {year}
  {2010})}\BibitemShut {NoStop}%
\bibitem [{\citenamefont {Leroux}\ \emph {et~al.}(2010)\citenamefont {Leroux},
  \citenamefont {Schleier-Smith},\ and\ \citenamefont {Vuletic}}]{Leroux2010}%
  \BibitemOpen
  \bibfield  {author} {\bibinfo {author} {\bibfnamefont {I.~D.}\ \bibnamefont
  {Leroux}}, \bibinfo {author} {\bibfnamefont {M.~H.}\ \bibnamefont
  {Schleier-Smith}},\ and\ \bibinfo {author} {\bibfnamefont {V.}~\bibnamefont
  {Vuletic}},\ }\bibfield  {title} {\bibinfo {title} {Implementation of cavity
  squeezing of a collective atomic spin},\ }\href
  {https://doi.org/10.1103/PhysRevLett.104.073602} {\bibfield  {journal}
  {\bibinfo  {journal} {Phys. Rev. Lett.}\ }\textbf {\bibinfo {volume} {104}},\
  \bibinfo {pages} {073602} (\bibinfo {year} {2010})}\BibitemShut {NoStop}%
\bibitem [{\citenamefont {Louchet-Chauvet}\ \emph {et~al.}(2010)\citenamefont
  {Louchet-Chauvet}, \citenamefont {Appel}, \citenamefont {Renema},
  \citenamefont {Oblak}, \citenamefont {Kjaergaard},\ and\ \citenamefont
  {Polzik}}]{Louchet-Chauvet2010}%
  \BibitemOpen
  \bibfield  {author} {\bibinfo {author} {\bibfnamefont {A.}~\bibnamefont
  {Louchet-Chauvet}}, \bibinfo {author} {\bibfnamefont {J.}~\bibnamefont
  {Appel}}, \bibinfo {author} {\bibfnamefont {J.~J.}\ \bibnamefont {Renema}},
  \bibinfo {author} {\bibfnamefont {D.}~\bibnamefont {Oblak}}, \bibinfo
  {author} {\bibfnamefont {N.}~\bibnamefont {Kjaergaard}},\ and\ \bibinfo
  {author} {\bibfnamefont {E.~S.}\ \bibnamefont {Polzik}},\ }\bibfield  {title}
  {\bibinfo {title} {Entanglement-assisted atomic clock beyond the projection
  noise limit},\ }\href {https://doi.org/10.1088/1367-2630/12/6/065032}
  {\bibfield  {journal} {\bibinfo  {journal} {New J. Phys.}\ }\textbf {\bibinfo
  {volume} {12}},\ \bibinfo {pages} {065032} (\bibinfo {year}
  {2010})}\BibitemShut {NoStop}%
\bibitem [{\citenamefont {Haas}\ \emph {et~al.}(2014)\citenamefont {Haas},
  \citenamefont {Volz}, \citenamefont {Gehr}, \citenamefont {Reichel},\ and\
  \citenamefont {Est{\`e}ve}}]{Haas2014}%
  \BibitemOpen
  \bibfield  {author} {\bibinfo {author} {\bibfnamefont {F.}~\bibnamefont
  {Haas}}, \bibinfo {author} {\bibfnamefont {J.}~\bibnamefont {Volz}}, \bibinfo
  {author} {\bibfnamefont {R.}~\bibnamefont {Gehr}}, \bibinfo {author}
  {\bibfnamefont {J.}~\bibnamefont {Reichel}},\ and\ \bibinfo {author}
  {\bibfnamefont {J.}~\bibnamefont {Est{\`e}ve}},\ }\bibfield  {title}
  {\bibinfo {title} {Entangled states of more than 40 atoms in an optical fiber
  cavity},\ }\href {https://doi.org/10.1126/science.1248905} {\bibfield
  {journal} {\bibinfo  {journal} {Science}\ }\textbf {\bibinfo {volume}
  {344}},\ \bibinfo {pages} {180} (\bibinfo {year} {2014})}\BibitemShut
  {NoStop}%
\bibitem [{\citenamefont {Hosten}\ \emph {et~al.}(2016)\citenamefont {Hosten},
  \citenamefont {Engelsen}, \citenamefont {Krishnakumar},\ and\ \citenamefont
  {Kasevich}}]{Hosten2016}%
  \BibitemOpen
  \bibfield  {author} {\bibinfo {author} {\bibfnamefont {O.}~\bibnamefont
  {Hosten}}, \bibinfo {author} {\bibfnamefont {N.~J.}\ \bibnamefont
  {Engelsen}}, \bibinfo {author} {\bibfnamefont {R.}~\bibnamefont
  {Krishnakumar}},\ and\ \bibinfo {author} {\bibfnamefont {M.~A.}\ \bibnamefont
  {Kasevich}},\ }\bibfield  {title} {\bibinfo {title} {Measurement noise 100
  times lower than the quantum-projection limit using entangled atoms},\ }\href
  {https://doi.org/10.1038/nature16176} {\bibfield  {journal} {\bibinfo
  {journal} {Nature}\ }\textbf {\bibinfo {volume} {529}},\ \bibinfo {pages}
  {505} (\bibinfo {year} {2016})}\BibitemShut {NoStop}%
\bibitem [{\citenamefont {Gross}\ \emph {et~al.}(2010)\citenamefont {Gross},
  \citenamefont {Zibold}, \citenamefont {Nicklas}, \citenamefont {Est\`{e}ve},\
  and\ \citenamefont {Oberthaler}}]{Gross2010}%
  \BibitemOpen
  \bibfield  {author} {\bibinfo {author} {\bibfnamefont {C.}~\bibnamefont
  {Gross}}, \bibinfo {author} {\bibfnamefont {T.}~\bibnamefont {Zibold}},
  \bibinfo {author} {\bibfnamefont {E.}~\bibnamefont {Nicklas}}, \bibinfo
  {author} {\bibfnamefont {J.}~\bibnamefont {Est\`{e}ve}},\ and\ \bibinfo
  {author} {\bibfnamefont {M.~K.}\ \bibnamefont {Oberthaler}},\ }\bibfield
  {title} {\bibinfo {title} {Nonlinear atom interferometer surpasses classical
  precision limit},\ }\href {https://doi.org/10.1038/nature08919} {\bibfield
  {journal} {\bibinfo  {journal} {Nature}\ }\textbf {\bibinfo {volume} {464}},\
  \bibinfo {pages} {1165} (\bibinfo {year} {2010})}\BibitemShut {NoStop}%
\bibitem [{\citenamefont {Riedel}\ \emph {et~al.}(2010)\citenamefont {Riedel},
  \citenamefont {B\"ohi}, \citenamefont {Li}, \citenamefont {H\"ansch},
  \citenamefont {Sinatra},\ and\ \citenamefont {Treutlein}}]{Riedel2010}%
  \BibitemOpen
  \bibfield  {author} {\bibinfo {author} {\bibfnamefont {M.}~\bibnamefont
  {Riedel}}, \bibinfo {author} {\bibfnamefont {P.}~\bibnamefont {B\"ohi}},
  \bibinfo {author} {\bibfnamefont {Y.}~\bibnamefont {Li}}, \bibinfo {author}
  {\bibfnamefont {T.}~\bibnamefont {H\"ansch}}, \bibinfo {author}
  {\bibfnamefont {A.}~\bibnamefont {Sinatra}},\ and\ \bibinfo {author}
  {\bibfnamefont {P.}~\bibnamefont {Treutlein}},\ }\bibfield  {title} {\bibinfo
  {title} {Atom-chip-based generation of entanglement for quantum metrology},\
  }\href {https://doi.org/10.1038/nature08988} {\bibfield  {journal} {\bibinfo
  {journal} {Nature}\ }\textbf {\bibinfo {volume} {464}},\ \bibinfo {pages}
  {1170} (\bibinfo {year} {2010})}\BibitemShut {NoStop}%
\bibitem [{\citenamefont {L{\"u}cke}\ \emph {et~al.}(2011)\citenamefont
  {L{\"u}cke}, \citenamefont {Scherer}, \citenamefont {Kruse}, \citenamefont
  {Pezz{\'e}}, \citenamefont {Deuretzbacher}, \citenamefont {Hyllus},
  \citenamefont {Topic}, \citenamefont {Peise}, \citenamefont {Ertmer},
  \citenamefont {Arlt}, \citenamefont {Santos}, \citenamefont {Smerzi},\ and\
  \citenamefont {Klempt}}]{Lucke2011}%
  \BibitemOpen
  \bibfield  {author} {\bibinfo {author} {\bibfnamefont {B.}~\bibnamefont
  {L{\"u}cke}}, \bibinfo {author} {\bibfnamefont {M.}~\bibnamefont {Scherer}},
  \bibinfo {author} {\bibfnamefont {J.}~\bibnamefont {Kruse}}, \bibinfo
  {author} {\bibfnamefont {L.}~\bibnamefont {Pezz{\'e}}}, \bibinfo {author}
  {\bibfnamefont {F.}~\bibnamefont {Deuretzbacher}}, \bibinfo {author}
  {\bibfnamefont {P.}~\bibnamefont {Hyllus}}, \bibinfo {author} {\bibfnamefont
  {O.}~\bibnamefont {Topic}}, \bibinfo {author} {\bibfnamefont
  {J.}~\bibnamefont {Peise}}, \bibinfo {author} {\bibfnamefont
  {W.}~\bibnamefont {Ertmer}}, \bibinfo {author} {\bibfnamefont
  {J.}~\bibnamefont {Arlt}}, \bibinfo {author} {\bibfnamefont {L.}~\bibnamefont
  {Santos}}, \bibinfo {author} {\bibfnamefont {A.}~\bibnamefont {Smerzi}},\
  and\ \bibinfo {author} {\bibfnamefont {C.}~\bibnamefont {Klempt}},\
  }\bibfield  {title} {\bibinfo {title} {Twin matter waves for interferometry
  beyond the classical limit},\ }\href
  {https://doi.org/10.1126/science.1208798} {\bibfield  {journal} {\bibinfo
  {journal} {Science}\ }\textbf {\bibinfo {volume} {334}},\ \bibinfo {pages}
  {773} (\bibinfo {year} {2011})}\BibitemShut {NoStop}%
\bibitem [{\citenamefont {Ockeloen}\ \emph {et~al.}(2013)\citenamefont
  {Ockeloen}, \citenamefont {Schmied}, \citenamefont {Riedel},\ and\
  \citenamefont {Treutlein}}]{Ockeloen2013}%
  \BibitemOpen
  \bibfield  {author} {\bibinfo {author} {\bibfnamefont {C.~F.}\ \bibnamefont
  {Ockeloen}}, \bibinfo {author} {\bibfnamefont {R.}~\bibnamefont {Schmied}},
  \bibinfo {author} {\bibfnamefont {M.~F.}\ \bibnamefont {Riedel}},\ and\
  \bibinfo {author} {\bibfnamefont {P.}~\bibnamefont {Treutlein}},\ }\bibfield
  {title} {\bibinfo {title} {Quantum metrology with a scanning probe atom
  interferometer},\ }\href {https://doi.org/10.1103/PhysRevLett.111.143001}
  {\bibfield  {journal} {\bibinfo  {journal} {Phys. Rev. Lett.}\ }\textbf
  {\bibinfo {volume} {111}},\ \bibinfo {pages} {143001} (\bibinfo {year}
  {2013})}\BibitemShut {NoStop}%
\bibitem [{\citenamefont {Strobel}\ \emph {et~al.}(2014)\citenamefont
  {Strobel}, \citenamefont {Muessel}, \citenamefont {Linnemann}, \citenamefont
  {Zibold}, \citenamefont {Hume}, \citenamefont {Pezz{\`e}}, \citenamefont
  {Smerzi},\ and\ \citenamefont {Oberthaler}}]{Strobel2014}%
  \BibitemOpen
  \bibfield  {author} {\bibinfo {author} {\bibfnamefont {H.}~\bibnamefont
  {Strobel}}, \bibinfo {author} {\bibfnamefont {W.}~\bibnamefont {Muessel}},
  \bibinfo {author} {\bibfnamefont {D.}~\bibnamefont {Linnemann}}, \bibinfo
  {author} {\bibfnamefont {T.}~\bibnamefont {Zibold}}, \bibinfo {author}
  {\bibfnamefont {D.~B.}\ \bibnamefont {Hume}}, \bibinfo {author}
  {\bibfnamefont {L.}~\bibnamefont {Pezz{\`e}}}, \bibinfo {author}
  {\bibfnamefont {A.}~\bibnamefont {Smerzi}},\ and\ \bibinfo {author}
  {\bibfnamefont {M.~K.}\ \bibnamefont {Oberthaler}},\ }\bibfield  {title}
  {\bibinfo {title} {Fisher information and entanglement of non-gaussian spin
  states},\ }\href {https://doi.org/10.1126/science.1250147} {\bibfield
  {journal} {\bibinfo  {journal} {Science}\ }\textbf {\bibinfo {volume}
  {345}},\ \bibinfo {pages} {424} (\bibinfo {year} {2014})}\BibitemShut
  {NoStop}%
\bibitem [{\citenamefont {Muessel}\ \emph {et~al.}(2014)\citenamefont
  {Muessel}, \citenamefont {Strobel}, \citenamefont {Linnemann}, \citenamefont
  {Hume},\ and\ \citenamefont {Oberthaler}}]{Muessel2014}%
  \BibitemOpen
  \bibfield  {author} {\bibinfo {author} {\bibfnamefont {W.}~\bibnamefont
  {Muessel}}, \bibinfo {author} {\bibfnamefont {H.}~\bibnamefont {Strobel}},
  \bibinfo {author} {\bibfnamefont {D.}~\bibnamefont {Linnemann}}, \bibinfo
  {author} {\bibfnamefont {D.~B.}\ \bibnamefont {Hume}},\ and\ \bibinfo
  {author} {\bibfnamefont {M.~K.}\ \bibnamefont {Oberthaler}},\ }\bibfield
  {title} {\bibinfo {title} {Scalable spin squeezing for quantum-enhanced
  magnetometry with {Bose-Einstein} condensates},\ }\href
  {https://doi.org/10.1103/PhysRevLett.113.103004} {\bibfield  {journal}
  {\bibinfo  {journal} {Phys. Rev. Lett.}\ }\textbf {\bibinfo {volume} {113}},\
  \bibinfo {pages} {103004} (\bibinfo {year} {2014})}\BibitemShut {NoStop}%
\bibitem [{\citenamefont {Kruse}\ \emph {et~al.}(2016)\citenamefont {Kruse},
  \citenamefont {Lange}, \citenamefont {Peise}, \citenamefont {L\"ucke},
  \citenamefont {Pezz\`e}, \citenamefont {Arlt}, \citenamefont {Ertmer},
  \citenamefont {Lisdat}, \citenamefont {Santos}, \citenamefont {Smerzi},\ and\
  \citenamefont {Klempt}}]{Kruse2016}%
  \BibitemOpen
  \bibfield  {author} {\bibinfo {author} {\bibfnamefont {I.}~\bibnamefont
  {Kruse}}, \bibinfo {author} {\bibfnamefont {K.}~\bibnamefont {Lange}},
  \bibinfo {author} {\bibfnamefont {J.}~\bibnamefont {Peise}}, \bibinfo
  {author} {\bibfnamefont {B.}~\bibnamefont {L\"ucke}}, \bibinfo {author}
  {\bibfnamefont {L.}~\bibnamefont {Pezz\`e}}, \bibinfo {author} {\bibfnamefont
  {J.}~\bibnamefont {Arlt}}, \bibinfo {author} {\bibfnamefont {W.}~\bibnamefont
  {Ertmer}}, \bibinfo {author} {\bibfnamefont {C.}~\bibnamefont {Lisdat}},
  \bibinfo {author} {\bibfnamefont {L.}~\bibnamefont {Santos}}, \bibinfo
  {author} {\bibfnamefont {A.}~\bibnamefont {Smerzi}},\ and\ \bibinfo {author}
  {\bibfnamefont {C.}~\bibnamefont {Klempt}},\ }\bibfield  {title} {\bibinfo
  {title} {Improvement of an atomic clock using squeezed vacuum},\ }\href
  {https://doi.org/10.1103/PhysRevLett.117.143004} {\bibfield  {journal}
  {\bibinfo  {journal} {Phys. Rev. Lett.}\ }\textbf {\bibinfo {volume} {117}},\
  \bibinfo {pages} {143004} (\bibinfo {year} {2016})}\BibitemShut {NoStop}%
\bibitem [{\citenamefont {Zou}\ \emph {et~al.}(2018)\citenamefont {Zou},
  \citenamefont {Wu}, \citenamefont {Liu}, \citenamefont {Luo}, \citenamefont
  {Guo}, \citenamefont {Cao}, \citenamefont {Tey},\ and\ \citenamefont
  {You}}]{Zou2018}%
  \BibitemOpen
  \bibfield  {author} {\bibinfo {author} {\bibfnamefont {Y.-Q.}\ \bibnamefont
  {Zou}}, \bibinfo {author} {\bibfnamefont {L.-N.}\ \bibnamefont {Wu}},
  \bibinfo {author} {\bibfnamefont {Q.}~\bibnamefont {Liu}}, \bibinfo {author}
  {\bibfnamefont {X.-Y.}\ \bibnamefont {Luo}}, \bibinfo {author} {\bibfnamefont
  {S.-F.}\ \bibnamefont {Guo}}, \bibinfo {author} {\bibfnamefont {J.-H.}\
  \bibnamefont {Cao}}, \bibinfo {author} {\bibfnamefont {M.~K.}\ \bibnamefont
  {Tey}},\ and\ \bibinfo {author} {\bibfnamefont {L.}~\bibnamefont {You}},\
  }\bibfield  {title} {\bibinfo {title} {Beating the classical precision limit
  with spin-1 {{Dicke}} states of more than 10,000 atoms},\ }\href
  {https://doi.org/10.1073/pnas.1715105115} {\bibfield  {journal} {\bibinfo
  {journal} {Proc. Natl. Acad. Sci. U.S.A.}\ }\textbf {\bibinfo {volume}
  {115}},\ \bibinfo {pages} {6381} (\bibinfo {year} {2018})}\BibitemShut
  {NoStop}%
\bibitem [{\citenamefont {Hyllus}\ \emph {et~al.}(2012)\citenamefont {Hyllus},
  \citenamefont {Pezz\'e}, \citenamefont {Smerzi},\ and\ \citenamefont
  {T\'oth}}]{Hyllus2012}%
  \BibitemOpen
  \bibfield  {author} {\bibinfo {author} {\bibfnamefont {P.}~\bibnamefont
  {Hyllus}}, \bibinfo {author} {\bibfnamefont {L.}~\bibnamefont {Pezz\'e}},
  \bibinfo {author} {\bibfnamefont {A.}~\bibnamefont {Smerzi}},\ and\ \bibinfo
  {author} {\bibfnamefont {G.}~\bibnamefont {T\'oth}},\ }\bibfield  {title}
  {\bibinfo {title} {Entanglement and extreme spin squeezing for a fluctuating
  number of indistinguishable particles},\ }\href
  {https://doi.org/10.1103/PhysRevA.86.012337} {\bibfield  {journal} {\bibinfo
  {journal} {Phys. Rev. A}\ }\textbf {\bibinfo {volume} {86}},\ \bibinfo
  {pages} {012337} (\bibinfo {year} {2012})}\BibitemShut {NoStop}%
\bibitem [{\citenamefont {L\"ucke}\ \emph {et~al.}(2014)\citenamefont
  {L\"ucke}, \citenamefont {Peise}, \citenamefont {Vitagliano}, \citenamefont
  {Arlt}, \citenamefont {Santos}, \citenamefont {T\'oth},\ and\ \citenamefont
  {Klempt}}]{Lucke2014}%
  \BibitemOpen
  \bibfield  {author} {\bibinfo {author} {\bibfnamefont {B.}~\bibnamefont
  {L\"ucke}}, \bibinfo {author} {\bibfnamefont {J.}~\bibnamefont {Peise}},
  \bibinfo {author} {\bibfnamefont {G.}~\bibnamefont {Vitagliano}}, \bibinfo
  {author} {\bibfnamefont {J.}~\bibnamefont {Arlt}}, \bibinfo {author}
  {\bibfnamefont {L.}~\bibnamefont {Santos}}, \bibinfo {author} {\bibfnamefont
  {G.}~\bibnamefont {T\'oth}},\ and\ \bibinfo {author} {\bibfnamefont
  {C.}~\bibnamefont {Klempt}},\ }\bibfield  {title} {\bibinfo {title}
  {Detecting multiparticle entanglement of {Dicke} states},\ }\href
  {https://doi.org/10.1103/PhysRevLett.112.155304} {\bibfield  {journal}
  {\bibinfo  {journal} {Phys. Rev. Lett.}\ }\textbf {\bibinfo {volume} {112}},\
  \bibinfo {pages} {155304} (\bibinfo {year} {2014})}\BibitemShut {NoStop}%
\bibitem [{\citenamefont {Est\`{e}ve}\ \emph {et~al.}(2008)\citenamefont
  {Est\`{e}ve}, \citenamefont {Gross}, \citenamefont {Weller}, \citenamefont
  {Giovanazzi},\ and\ \citenamefont {Oberthaler}}]{Esteve2008}%
  \BibitemOpen
  \bibfield  {author} {\bibinfo {author} {\bibfnamefont {J.}~\bibnamefont
  {Est\`{e}ve}}, \bibinfo {author} {\bibfnamefont {C.}~\bibnamefont {Gross}},
  \bibinfo {author} {\bibfnamefont {A.}~\bibnamefont {Weller}}, \bibinfo
  {author} {\bibfnamefont {S.}~\bibnamefont {Giovanazzi}},\ and\ \bibinfo
  {author} {\bibfnamefont {M.~K.}\ \bibnamefont {Oberthaler}},\ }\bibfield
  {title} {\bibinfo {title} {Squeezing and entanglement in a {Bose-Einstein}
  condensate},\ }\href {https://doi.org/10.1038/nature07332} {\bibfield
  {journal} {\bibinfo  {journal} {Nature}\ }\textbf {\bibinfo {volume} {455}},\
  \bibinfo {pages} {1216} (\bibinfo {year} {2008})}\BibitemShut {NoStop}%
\bibitem [{\citenamefont {Berrada}\ \emph {et~al.}(2013)\citenamefont
  {Berrada}, \citenamefont {van Frank}, \citenamefont {B\"ucker}, \citenamefont
  {Schumm}, \citenamefont {Schaff},\ and\ \citenamefont
  {Schmiedmayer}}]{Berrada2013}%
  \BibitemOpen
  \bibfield  {author} {\bibinfo {author} {\bibfnamefont {T.}~\bibnamefont
  {Berrada}}, \bibinfo {author} {\bibfnamefont {S.}~\bibnamefont {van Frank}},
  \bibinfo {author} {\bibfnamefont {R.}~\bibnamefont {B\"ucker}}, \bibinfo
  {author} {\bibfnamefont {T.}~\bibnamefont {Schumm}}, \bibinfo {author}
  {\bibfnamefont {J.-F.}\ \bibnamefont {Schaff}},\ and\ \bibinfo {author}
  {\bibfnamefont {J.}~\bibnamefont {Schmiedmayer}},\ }\bibfield  {title}
  {\bibinfo {title} {Integrated {Mach-Zehnder} interferometer for
  {Bose-Einstein} condensates},\ }\href {https://doi.org/10.1038/ncomms3077}
  {\bibfield  {journal} {\bibinfo  {journal} {Nat. Commun.}\ }\textbf {\bibinfo
  {volume} {4}},\  (\bibinfo {year} {2013})}\BibitemShut {NoStop}%
\bibitem [{\citenamefont {Lange}\ \emph {et~al.}(2018)\citenamefont {Lange},
  \citenamefont {Peise}, \citenamefont {L{\"u}cke}, \citenamefont {Kruse},
  \citenamefont {Vitagliano}, \citenamefont {Apellaniz}, \citenamefont
  {Kleinmann}, \citenamefont {T\'oth},\ and\ \citenamefont
  {Klempt}}]{Lange2018}%
  \BibitemOpen
  \bibfield  {author} {\bibinfo {author} {\bibfnamefont {K.}~\bibnamefont
  {Lange}}, \bibinfo {author} {\bibfnamefont {J.}~\bibnamefont {Peise}},
  \bibinfo {author} {\bibfnamefont {B.}~\bibnamefont {L{\"u}cke}}, \bibinfo
  {author} {\bibfnamefont {I.}~\bibnamefont {Kruse}}, \bibinfo {author}
  {\bibfnamefont {G.}~\bibnamefont {Vitagliano}}, \bibinfo {author}
  {\bibfnamefont {I.}~\bibnamefont {Apellaniz}}, \bibinfo {author}
  {\bibfnamefont {M.}~\bibnamefont {Kleinmann}}, \bibinfo {author}
  {\bibfnamefont {G.}~\bibnamefont {T\'oth}},\ and\ \bibinfo {author}
  {\bibfnamefont {C.}~\bibnamefont {Klempt}},\ }\bibfield  {title} {\bibinfo
  {title} {Entanglement between two spatially separated atomic modes},\ }\href
  {https://doi.org/10.1126/science.aao2035} {\bibfield  {journal} {\bibinfo
  {journal} {Science}\ }\textbf {\bibinfo {volume} {360}},\ \bibinfo {pages}
  {416} (\bibinfo {year} {2018})}\BibitemShut {NoStop}%
\bibitem [{\citenamefont {Fadel}\ \emph {et~al.}(2018)\citenamefont {Fadel},
  \citenamefont {Zibold}, \citenamefont {D{\'e}camps},\ and\ \citenamefont
  {Treutlein}}]{Fadel2018}%
  \BibitemOpen
  \bibfield  {author} {\bibinfo {author} {\bibfnamefont {M.}~\bibnamefont
  {Fadel}}, \bibinfo {author} {\bibfnamefont {T.}~\bibnamefont {Zibold}},
  \bibinfo {author} {\bibfnamefont {B.}~\bibnamefont {D{\'e}camps}},\ and\
  \bibinfo {author} {\bibfnamefont {P.}~\bibnamefont {Treutlein}},\ }\bibfield
  {title} {\bibinfo {title} {Spatial entanglement patterns and
  {Einstein-Podolsky-Rosen} steering in {Bose-Einstein} condensates},\ }\href
  {https://doi.org/10.1126/science.aao1850} {\bibfield  {journal} {\bibinfo
  {journal} {Science}\ }\textbf {\bibinfo {volume} {360}},\ \bibinfo {pages}
  {409} (\bibinfo {year} {2018})}\BibitemShut {NoStop}%
\bibitem [{\citenamefont {Kunkel}\ \emph {et~al.}(2018)\citenamefont {Kunkel},
  \citenamefont {Pr{\"u}fer}, \citenamefont {Strobel}, \citenamefont
  {Linnemann}, \citenamefont {Fr{\"o}lian}, \citenamefont {Gasenzer},
  \citenamefont {G{\"a}rttner},\ and\ \citenamefont {Oberthaler}}]{Kunkel2018}%
  \BibitemOpen
  \bibfield  {author} {\bibinfo {author} {\bibfnamefont {P.}~\bibnamefont
  {Kunkel}}, \bibinfo {author} {\bibfnamefont {M.}~\bibnamefont {Pr{\"u}fer}},
  \bibinfo {author} {\bibfnamefont {H.}~\bibnamefont {Strobel}}, \bibinfo
  {author} {\bibfnamefont {D.}~\bibnamefont {Linnemann}}, \bibinfo {author}
  {\bibfnamefont {A.}~\bibnamefont {Fr{\"o}lian}}, \bibinfo {author}
  {\bibfnamefont {T.}~\bibnamefont {Gasenzer}}, \bibinfo {author}
  {\bibfnamefont {M.}~\bibnamefont {G{\"a}rttner}},\ and\ \bibinfo {author}
  {\bibfnamefont {M.~K.}\ \bibnamefont {Oberthaler}},\ }\bibfield  {title}
  {\bibinfo {title} {Spatially distributed multipartite entanglement enables
  {EPR} steering of atomic clouds},\ }\href
  {https://doi.org/10.1126/science.aao2254} {\bibfield  {journal} {\bibinfo
  {journal} {Science}\ }\textbf {\bibinfo {volume} {360}},\ \bibinfo {pages}
  {413} (\bibinfo {year} {2018})}\BibitemShut {NoStop}%
\bibitem [{\citenamefont {Bucker}\ \emph {et~al.}(2011)\citenamefont {Bucker},
  \citenamefont {Grond}, \citenamefont {Manz}, \citenamefont {Berrada},
  \citenamefont {Betz}, \citenamefont {Koller}, \citenamefont {Hohenester},
  \citenamefont {Schumm}, \citenamefont {Perrin},\ and\ \citenamefont
  {Schmiedmayer}}]{Bucker2011}%
  \BibitemOpen
  \bibfield  {author} {\bibinfo {author} {\bibfnamefont {R.}~\bibnamefont
  {Bucker}}, \bibinfo {author} {\bibfnamefont {J.}~\bibnamefont {Grond}},
  \bibinfo {author} {\bibfnamefont {S.}~\bibnamefont {Manz}}, \bibinfo {author}
  {\bibfnamefont {T.}~\bibnamefont {Berrada}}, \bibinfo {author} {\bibfnamefont
  {T.}~\bibnamefont {Betz}}, \bibinfo {author} {\bibfnamefont {C.}~\bibnamefont
  {Koller}}, \bibinfo {author} {\bibfnamefont {U.}~\bibnamefont {Hohenester}},
  \bibinfo {author} {\bibfnamefont {T.}~\bibnamefont {Schumm}}, \bibinfo
  {author} {\bibfnamefont {A.}~\bibnamefont {Perrin}},\ and\ \bibinfo {author}
  {\bibfnamefont {J.}~\bibnamefont {Schmiedmayer}},\ }\bibfield  {title}
  {\bibinfo {title} {Twin-atom beams},\ }\href
  {https://doi.org/10.1038/nphys1992} {\bibfield  {journal} {\bibinfo
  {journal} {Nature Phys.}\ }\textbf {\bibinfo {volume} {7}},\ \bibinfo {pages}
  {608} (\bibinfo {year} {2011})}\BibitemShut {NoStop}%
\bibitem [{\citenamefont {Kheruntsyan}\ \emph {et~al.}(2012)\citenamefont
  {Kheruntsyan}, \citenamefont {Jaskula}, \citenamefont {Deuar}, \citenamefont
  {Bonneau}, \citenamefont {Partridge}, \citenamefont {Ruaudel}, \citenamefont
  {Lopes}, \citenamefont {Boiron},\ and\ \citenamefont
  {Westbrook}}]{Kheruntsyan2012}%
  \BibitemOpen
  \bibfield  {author} {\bibinfo {author} {\bibfnamefont {K.~V.}\ \bibnamefont
  {Kheruntsyan}}, \bibinfo {author} {\bibfnamefont {J.-C.}\ \bibnamefont
  {Jaskula}}, \bibinfo {author} {\bibfnamefont {P.}~\bibnamefont {Deuar}},
  \bibinfo {author} {\bibfnamefont {M.}~\bibnamefont {Bonneau}}, \bibinfo
  {author} {\bibfnamefont {G.~B.}\ \bibnamefont {Partridge}}, \bibinfo {author}
  {\bibfnamefont {J.}~\bibnamefont {Ruaudel}}, \bibinfo {author} {\bibfnamefont
  {R.}~\bibnamefont {Lopes}}, \bibinfo {author} {\bibfnamefont
  {D.}~\bibnamefont {Boiron}},\ and\ \bibinfo {author} {\bibfnamefont {C.~I.}\
  \bibnamefont {Westbrook}},\ }\bibfield  {title} {\bibinfo {title} {Violation
  of the {Cauchy}-{Schwarz} {Inequality} with {Matter} {Waves}},\ }\href
  {https://doi.org/10.1103/PhysRevLett.108.260401} {\bibfield  {journal}
  {\bibinfo  {journal} {Phys. Rev. Lett.}\ }\textbf {\bibinfo {volume} {108}},\
  \bibinfo {pages} {260401} (\bibinfo {year} {2012})}\BibitemShut {NoStop}%
\bibitem [{\citenamefont {Shin}\ \emph {et~al.}(2019)\citenamefont {Shin},
  \citenamefont {Henson}, \citenamefont {Hodgman}, \citenamefont {Wasak},
  \citenamefont {Chwede{\'{n}}czuk},\ and\ \citenamefont
  {Truscott}}]{Shin2019}%
  \BibitemOpen
  \bibfield  {author} {\bibinfo {author} {\bibfnamefont {D.~K.}\ \bibnamefont
  {Shin}}, \bibinfo {author} {\bibfnamefont {B.~M.}\ \bibnamefont {Henson}},
  \bibinfo {author} {\bibfnamefont {S.~S.}\ \bibnamefont {Hodgman}}, \bibinfo
  {author} {\bibfnamefont {T.}~\bibnamefont {Wasak}}, \bibinfo {author}
  {\bibfnamefont {J.}~\bibnamefont {Chwede{\'{n}}czuk}},\ and\ \bibinfo
  {author} {\bibfnamefont {A.~G.}\ \bibnamefont {Truscott}},\ }\bibfield
  {title} {\bibinfo {title} {Bell correlations between spatially separated
  pairs of atoms},\ }\href {https://doi.org/10.1038/s41467-019-12192-8}
  {\bibfield  {journal} {\bibinfo  {journal} {Nature Communications}\ }\textbf
  {\bibinfo {volume} {10}},\ \bibinfo {pages} {4447} (\bibinfo {year}
  {2019})}\BibitemShut {NoStop}%
\bibitem [{\citenamefont {Salvi}\ \emph {et~al.}(2018)\citenamefont {Salvi},
  \citenamefont {Poli}, \citenamefont {Vuleti{\'{c}}},\ and\ \citenamefont
  {Tino}}]{Salvi2018}%
  \BibitemOpen
  \bibfield  {author} {\bibinfo {author} {\bibfnamefont {L.}~\bibnamefont
  {Salvi}}, \bibinfo {author} {\bibfnamefont {N.}~\bibnamefont {Poli}},
  \bibinfo {author} {\bibfnamefont {V.}~\bibnamefont {Vuleti{\'{c}}}},\ and\
  \bibinfo {author} {\bibfnamefont {G.~M.}\ \bibnamefont {Tino}},\ }\bibfield
  {title} {\bibinfo {title} {Squeezing on momentum states for atom
  interferometry},\ }\href {https://doi.org/10.1103/physrevlett.120.033601}
  {\bibfield  {journal} {\bibinfo  {journal} {Phys. Rev. Lett.}\ }\textbf
  {\bibinfo {volume} {120}},\ \bibinfo {pages} {033601} (\bibinfo {year}
  {2018})}\BibitemShut {NoStop}%
\bibitem [{\citenamefont {Geiger}\ and\ \citenamefont
  {Trupke}(2018)}]{Geiger2018}%
  \BibitemOpen
  \bibfield  {author} {\bibinfo {author} {\bibfnamefont {R.}~\bibnamefont
  {Geiger}}\ and\ \bibinfo {author} {\bibfnamefont {M.}~\bibnamefont
  {Trupke}},\ }\bibfield  {title} {\bibinfo {title} {Proposal for a quantum
  test of the weak equivalence principle with entangled atomic species},\
  }\href {https://doi.org/10.1103/PhysRevLett.120.043602} {\bibfield  {journal}
  {\bibinfo  {journal} {Phys. Rev. Lett.}\ }\textbf {\bibinfo {volume} {120}},\
  \bibinfo {pages} {043602} (\bibinfo {year} {2018})}\BibitemShut {NoStop}%
\bibitem [{\citenamefont {Shankar}\ \emph {et~al.}(2019)\citenamefont
  {Shankar}, \citenamefont {Salvi}, \citenamefont {Chiofalo}, \citenamefont
  {Poli},\ and\ \citenamefont {Holland}}]{Shankar2019}%
  \BibitemOpen
  \bibfield  {author} {\bibinfo {author} {\bibfnamefont {A.}~\bibnamefont
  {Shankar}}, \bibinfo {author} {\bibfnamefont {L.}~\bibnamefont {Salvi}},
  \bibinfo {author} {\bibfnamefont {M.~L.}\ \bibnamefont {Chiofalo}}, \bibinfo
  {author} {\bibfnamefont {N.}~\bibnamefont {Poli}},\ and\ \bibinfo {author}
  {\bibfnamefont {M.~J.}\ \bibnamefont {Holland}},\ }\bibfield  {title}
  {\bibinfo {title} {Squeezed state metrology with bragg interferometers
  operating in a cavity},\ }\href {https://doi.org/10.1088/2058-9565/ab455d}
  {\bibfield  {journal} {\bibinfo  {journal} {Quantum Science and Technology}\
  }\textbf {\bibinfo {volume} {4}},\ \bibinfo {pages} {045010} (\bibinfo {year}
  {2019})}\BibitemShut {NoStop}%
\bibitem [{\citenamefont {Szigeti}\ \emph
  {et~al.}(2020{\natexlab{a}})\citenamefont {Szigeti}, \citenamefont {Nolan},
  \citenamefont {Close},\ and\ \citenamefont {Haine}}]{Szigeti2020}%
  \BibitemOpen
  \bibfield  {author} {\bibinfo {author} {\bibfnamefont {S.~S.}\ \bibnamefont
  {Szigeti}}, \bibinfo {author} {\bibfnamefont {S.~P.}\ \bibnamefont {Nolan}},
  \bibinfo {author} {\bibfnamefont {J.~D.}\ \bibnamefont {Close}},\ and\
  \bibinfo {author} {\bibfnamefont {S.~A.}\ \bibnamefont {Haine}},\ }\bibfield
  {title} {\bibinfo {title} {High-precision quantum-enhanced gravimetry with a
  bose-einstein condensate},\ }\href
  {https://doi.org/10.1103/PhysRevLett.125.100402} {\bibfield  {journal}
  {\bibinfo  {journal} {Phys. Rev. Lett.}\ }\textbf {\bibinfo {volume} {125}},\
  \bibinfo {pages} {100402} (\bibinfo {year} {2020}{\natexlab{a}})}\BibitemShut
  {NoStop}%
\bibitem [{\citenamefont {Szigeti}\ \emph
  {et~al.}(2020{\natexlab{b}})\citenamefont {Szigeti}, \citenamefont {Hosten},\
  and\ \citenamefont {Haine}}]{Szigeti2020a}%
  \BibitemOpen
  \bibfield  {author} {\bibinfo {author} {\bibfnamefont {S.~S.}\ \bibnamefont
  {Szigeti}}, \bibinfo {author} {\bibfnamefont {O.}~\bibnamefont {Hosten}},\
  and\ \bibinfo {author} {\bibfnamefont {S.~A.}\ \bibnamefont {Haine}},\
  }\bibfield  {title} {\bibinfo {title} {Will quantum-enhanced atom
  interferometry ever be useful? prospects for improving cold-atom sensors with
  quantum entanglement},\ }\href@noop {} {\bibfield  {journal} {\bibinfo
  {journal} {arXiv:2010.09168}\ } (\bibinfo {year} {2020}{\natexlab{b}})},\
  \Eprint {https://arxiv.org/abs/2010.09168} {arXiv:2010.09168 [quant-ph]}
  \BibitemShut {NoStop}%
\bibitem [{\citenamefont {Snadden}\ \emph {et~al.}(1998)\citenamefont
  {Snadden}, \citenamefont {McGuirk}, \citenamefont {Bouyer}, \citenamefont
  {Haritos},\ and\ \citenamefont {Kasevich}}]{Snadden1998}%
  \BibitemOpen
  \bibfield  {author} {\bibinfo {author} {\bibfnamefont {M.}~\bibnamefont
  {Snadden}}, \bibinfo {author} {\bibfnamefont {J.}~\bibnamefont {McGuirk}},
  \bibinfo {author} {\bibfnamefont {P.}~\bibnamefont {Bouyer}}, \bibinfo
  {author} {\bibfnamefont {K.}~\bibnamefont {Haritos}},\ and\ \bibinfo {author}
  {\bibfnamefont {M.}~\bibnamefont {Kasevich}},\ }\bibfield  {title} {\bibinfo
  {title} {Measurement of the earth's gravity gradient with an atom
  interferometer-based gravity gradiometer},\ }\href
  {https://doi.org/10.1103/physrevlett.81.971} {\bibfield  {journal} {\bibinfo
  {journal} {Physical Review Letters}\ }\textbf {\bibinfo {volume} {81}},\
  \bibinfo {pages} {971} (\bibinfo {year} {1998})}\BibitemShut {NoStop}%
\bibitem [{\citenamefont {Aguilera}\ \emph {et~al.}(2014)\citenamefont
  {Aguilera}, \citenamefont {Ahlers}, \citenamefont {Battelier}, \citenamefont
  {Bawamia}, \citenamefont {Bertoldi}, \citenamefont {Bondarescu},
  \citenamefont {Bongs}, \citenamefont {Bouyer}, \citenamefont {Braxmaier},
  \citenamefont {{L Cacciapuoti}}, \citenamefont {Chaloner}, \citenamefont
  {Chwalla}, \citenamefont {Ertmer}, \citenamefont {Franz}, \citenamefont
  {Gaaloul} \emph {et~al.}}]{Aguilera2014a}%
  \BibitemOpen
  \bibfield  {author} {\bibinfo {author} {\bibfnamefont {D.~N.}\ \bibnamefont
  {Aguilera}}, \bibinfo {author} {\bibfnamefont {H.}~\bibnamefont {Ahlers}},
  \bibinfo {author} {\bibfnamefont {B.}~\bibnamefont {Battelier}}, \bibinfo
  {author} {\bibfnamefont {A.}~\bibnamefont {Bawamia}}, \bibinfo {author}
  {\bibfnamefont {A.}~\bibnamefont {Bertoldi}}, \bibinfo {author}
  {\bibfnamefont {R.}~\bibnamefont {Bondarescu}}, \bibinfo {author}
  {\bibfnamefont {K.}~\bibnamefont {Bongs}}, \bibinfo {author} {\bibfnamefont
  {P.}~\bibnamefont {Bouyer}}, \bibinfo {author} {\bibfnamefont
  {C.}~\bibnamefont {Braxmaier}}, \bibinfo {author} {\bibnamefont {{L
  Cacciapuoti}}}, \bibinfo {author} {\bibfnamefont {C.}~\bibnamefont
  {Chaloner}}, \bibinfo {author} {\bibfnamefont {M.}~\bibnamefont {Chwalla}},
  \bibinfo {author} {\bibfnamefont {W.}~\bibnamefont {Ertmer}}, \bibinfo
  {author} {\bibfnamefont {M.}~\bibnamefont {Franz}}, \bibinfo {author}
  {\bibfnamefont {N.}~\bibnamefont {Gaaloul}}, \emph {et~al.},\ }\bibfield
  {title} {\bibinfo {title} {{STE}-{QUEST}{\textemdash}test of the universality
  of free fall using cold atom interferometry},\ }\href
  {https://doi.org/10.1088/0264-9381/31/11/115010} {\bibfield  {journal}
  {\bibinfo  {journal} {Class. Quant. Grav.}\ }\textbf {\bibinfo {volume}
  {31}},\ \bibinfo {pages} {115010} (\bibinfo {year} {2014})}\BibitemShut
  {NoStop}%
\bibitem [{\citenamefont {Tino}\ \emph {et~al.}(2019)\citenamefont {Tino},
  \citenamefont {Bassi}, \citenamefont {Bianco}, \citenamefont {Bongs},
  \citenamefont {Bouyer}, \citenamefont {Cacciapuoti}, \citenamefont
  {Capozziello}, \citenamefont {Chen}, \citenamefont {Chiofalo}, \citenamefont
  {Derevianko}, \citenamefont {Ertmer}, \citenamefont {Gaaloul}, \citenamefont
  {Gill}, \citenamefont {Graham}, \citenamefont {Hogan} \emph
  {et~al.}}]{Tino2019}%
  \BibitemOpen
  \bibfield  {author} {\bibinfo {author} {\bibfnamefont {G.~M.}\ \bibnamefont
  {Tino}}, \bibinfo {author} {\bibfnamefont {A.}~\bibnamefont {Bassi}},
  \bibinfo {author} {\bibfnamefont {G.}~\bibnamefont {Bianco}}, \bibinfo
  {author} {\bibfnamefont {K.}~\bibnamefont {Bongs}}, \bibinfo {author}
  {\bibfnamefont {P.}~\bibnamefont {Bouyer}}, \bibinfo {author} {\bibfnamefont
  {L.}~\bibnamefont {Cacciapuoti}}, \bibinfo {author} {\bibfnamefont
  {S.}~\bibnamefont {Capozziello}}, \bibinfo {author} {\bibfnamefont
  {X.}~\bibnamefont {Chen}}, \bibinfo {author} {\bibfnamefont {M.~L.}\
  \bibnamefont {Chiofalo}}, \bibinfo {author} {\bibfnamefont {A.}~\bibnamefont
  {Derevianko}}, \bibinfo {author} {\bibfnamefont {W.}~\bibnamefont {Ertmer}},
  \bibinfo {author} {\bibfnamefont {N.}~\bibnamefont {Gaaloul}}, \bibinfo
  {author} {\bibfnamefont {P.}~\bibnamefont {Gill}}, \bibinfo {author}
  {\bibfnamefont {P.~W.}\ \bibnamefont {Graham}}, \bibinfo {author}
  {\bibfnamefont {J.~M.}\ \bibnamefont {Hogan}}, \emph {et~al.},\ }\bibfield
  {title} {\bibinfo {title} {Sage: A proposal for a space atomic gravity
  explorer},\ }\href {https://doi.org/10.1140/epjd/e2019-100324-6} {\bibfield
  {journal} {\bibinfo  {journal} {The European Physical Journal D}\ }\textbf
  {\bibinfo {volume} {73}},\ \bibinfo {pages} {228} (\bibinfo {year}
  {2019})}\BibitemShut {NoStop}%
\bibitem [{\citenamefont {Canuel}\ \emph {et~al.}(2018)\citenamefont {Canuel},
  \citenamefont {Bertoldi}, \citenamefont {Amand}, \citenamefont {di~Borgo},
  \citenamefont {Chantrait}, \citenamefont {Danquigny}, \citenamefont
  {{\'{A}}lvarez}, \citenamefont {Fang}, \citenamefont {Freise}, \citenamefont
  {Geiger}, \citenamefont {Gillot}, \citenamefont {Henry}, \citenamefont
  {Hinderer}, \citenamefont {Holleville}, \citenamefont {Junca} \emph
  {et~al.}}]{Canuel2018}%
  \BibitemOpen
  \bibfield  {author} {\bibinfo {author} {\bibfnamefont {B.}~\bibnamefont
  {Canuel}}, \bibinfo {author} {\bibfnamefont {A.}~\bibnamefont {Bertoldi}},
  \bibinfo {author} {\bibfnamefont {L.}~\bibnamefont {Amand}}, \bibinfo
  {author} {\bibfnamefont {E.~P.}\ \bibnamefont {di~Borgo}}, \bibinfo {author}
  {\bibfnamefont {T.}~\bibnamefont {Chantrait}}, \bibinfo {author}
  {\bibfnamefont {C.}~\bibnamefont {Danquigny}}, \bibinfo {author}
  {\bibfnamefont {M.~D.}\ \bibnamefont {{\'{A}}lvarez}}, \bibinfo {author}
  {\bibfnamefont {B.}~\bibnamefont {Fang}}, \bibinfo {author} {\bibfnamefont
  {A.}~\bibnamefont {Freise}}, \bibinfo {author} {\bibfnamefont
  {R.}~\bibnamefont {Geiger}}, \bibinfo {author} {\bibfnamefont
  {J.}~\bibnamefont {Gillot}}, \bibinfo {author} {\bibfnamefont
  {S.}~\bibnamefont {Henry}}, \bibinfo {author} {\bibfnamefont
  {J.}~\bibnamefont {Hinderer}}, \bibinfo {author} {\bibfnamefont
  {D.}~\bibnamefont {Holleville}}, \bibinfo {author} {\bibfnamefont
  {J.}~\bibnamefont {Junca}}, \emph {et~al.},\ }\bibfield  {title} {\bibinfo
  {title} {Exploring gravity with the {MIGA} large scale atom interferometer},\
  }\bibfield  {journal} {\bibinfo  {journal} {Sci. Rep.}\ }\textbf {\bibinfo
  {volume} {8}},\ \href {https://doi.org/10.1038/s41598-018-32165-z}
  {10.1038/s41598-018-32165-z} (\bibinfo {year} {2018})\BibitemShut {NoStop}%
\bibitem [{\citenamefont {Canuel}\ \emph {et~al.}(2020)\citenamefont {Canuel},
  \citenamefont {Abend}, \citenamefont {Amaro-Seoane}, \citenamefont
  {Badaracco}, \citenamefont {Beaufils}, \citenamefont {Bertoldi},
  \citenamefont {Bongs}, \citenamefont {Bouyer}, \citenamefont {Braxmaier},
  \citenamefont {Chaibi}, \citenamefont {Christensen}, \citenamefont {Fitzek},
  \citenamefont {Flouris}, \citenamefont {Gaaloul}, \citenamefont {Gaffet}
  \emph {et~al.}}]{Canuel2020}%
  \BibitemOpen
  \bibfield  {author} {\bibinfo {author} {\bibfnamefont {B.}~\bibnamefont
  {Canuel}}, \bibinfo {author} {\bibfnamefont {S.}~\bibnamefont {Abend}},
  \bibinfo {author} {\bibfnamefont {P.}~\bibnamefont {Amaro-Seoane}}, \bibinfo
  {author} {\bibfnamefont {F.}~\bibnamefont {Badaracco}}, \bibinfo {author}
  {\bibfnamefont {Q.}~\bibnamefont {Beaufils}}, \bibinfo {author}
  {\bibfnamefont {A.}~\bibnamefont {Bertoldi}}, \bibinfo {author}
  {\bibfnamefont {K.}~\bibnamefont {Bongs}}, \bibinfo {author} {\bibfnamefont
  {P.}~\bibnamefont {Bouyer}}, \bibinfo {author} {\bibfnamefont
  {C.}~\bibnamefont {Braxmaier}}, \bibinfo {author} {\bibfnamefont
  {W.}~\bibnamefont {Chaibi}}, \bibinfo {author} {\bibfnamefont
  {N.}~\bibnamefont {Christensen}}, \bibinfo {author} {\bibfnamefont
  {F.}~\bibnamefont {Fitzek}}, \bibinfo {author} {\bibfnamefont
  {G.}~\bibnamefont {Flouris}}, \bibinfo {author} {\bibfnamefont
  {N.}~\bibnamefont {Gaaloul}}, \bibinfo {author} {\bibfnamefont
  {S.}~\bibnamefont {Gaffet}}, \emph {et~al.},\ }\bibfield  {title} {\bibinfo
  {title} {Elgar - a european laboratory for gravitation and
  atom-interferometric research},\ }\href
  {http://iopscience.iop.org/10.1088/1361-6382/aba80e} {\bibfield  {journal}
  {\bibinfo  {journal} {Classical and Quantum Gravity}\ } (\bibinfo {year}
  {2020})}\BibitemShut {NoStop}%
\bibitem [{\citenamefont {Schrinski}\ \emph {et~al.}(2020)\citenamefont
  {Schrinski}, \citenamefont {Hornberger},\ and\ \citenamefont
  {Nimmrichter}}]{Schrinski2020}%
  \BibitemOpen
  \bibfield  {author} {\bibinfo {author} {\bibfnamefont {B.}~\bibnamefont
  {Schrinski}}, \bibinfo {author} {\bibfnamefont {K.}~\bibnamefont
  {Hornberger}},\ and\ \bibinfo {author} {\bibfnamefont {S.}~\bibnamefont
  {Nimmrichter}},\ }\bibfield  {title} {\bibinfo {title} {How to rule out
  collapse models with {BEC} interferometry},\ }\href@noop {} {\bibfield
  {journal} {\bibinfo  {journal} {arXiv:2008.13580}\ } (\bibinfo {year}
  {2020})}\BibitemShut {NoStop}%
\bibitem [{\citenamefont {Lalo\"e}\ and\ \citenamefont
  {Mullin}(2009)}]{Laloe2009}%
  \BibitemOpen
  \bibfield  {author} {\bibinfo {author} {\bibfnamefont {F.}~\bibnamefont
  {Lalo\"e}}\ and\ \bibinfo {author} {\bibfnamefont {W.~J.}\ \bibnamefont
  {Mullin}},\ }\bibfield  {title} {\bibinfo {title} {Interferometry with
  independent {Bose-Einstein} condensates: parity as an {EPR}/{Bell} quantum
  variable},\ }\href {https://doi.org/10.1140/epjb/e2009-00248-6} {\bibfield
  {journal} {\bibinfo  {journal} {The European Physical Journal B}\ }\textbf
  {\bibinfo {volume} {70}},\ \bibinfo {pages} {377} (\bibinfo {year}
  {2009})}\BibitemShut {NoStop}%
\bibitem [{\citenamefont {L\"ucke}\ \emph {et~al.}(2011)\citenamefont
  {L\"ucke}, \citenamefont {Scherer}, \citenamefont {Kruse}, \citenamefont
  {Pezz\'e}, \citenamefont {Deuretzbacher}, \citenamefont {Hyllus},
  \citenamefont {Topic}, \citenamefont {Peise}, \citenamefont {Ertmer},
  \citenamefont {Arlt}, \citenamefont {Santos}, \citenamefont {Smerzi},\ and\
  \citenamefont {Klempt}}]{Luecke2011a}%
  \BibitemOpen
  \bibfield  {author} {\bibinfo {author} {\bibfnamefont {B.}~\bibnamefont
  {L\"ucke}}, \bibinfo {author} {\bibfnamefont {M.}~\bibnamefont {Scherer}},
  \bibinfo {author} {\bibfnamefont {J.}~\bibnamefont {Kruse}}, \bibinfo
  {author} {\bibfnamefont {L.}~\bibnamefont {Pezz\'e}}, \bibinfo {author}
  {\bibfnamefont {F.}~\bibnamefont {Deuretzbacher}}, \bibinfo {author}
  {\bibfnamefont {P.}~\bibnamefont {Hyllus}}, \bibinfo {author} {\bibfnamefont
  {O.}~\bibnamefont {Topic}}, \bibinfo {author} {\bibfnamefont
  {J.}~\bibnamefont {Peise}}, \bibinfo {author} {\bibfnamefont
  {W.}~\bibnamefont {Ertmer}}, \bibinfo {author} {\bibfnamefont
  {J.}~\bibnamefont {Arlt}}, \bibinfo {author} {\bibfnamefont {L.}~\bibnamefont
  {Santos}}, \bibinfo {author} {\bibfnamefont {A.}~\bibnamefont {Smerzi}},\
  and\ \bibinfo {author} {\bibfnamefont {C.}~\bibnamefont {Klempt}},\
  }\bibfield  {title} {\bibinfo {title} {Twin matter waves for interferometry
  beyond the classical limit},\ }\href
  {http://www.sciencemag.org/content/334/6057/773.abstract} {\bibfield
  {journal} {\bibinfo  {journal} {Science}\ }\textbf {\bibinfo {volume}
  {334}},\ \bibinfo {pages} {773} (\bibinfo {year} {2011})}\BibitemShut
  {NoStop}%
\bibitem [{\citenamefont {Gross}\ \emph {et~al.}(2011)\citenamefont {Gross},
  \citenamefont {Strobel}, \citenamefont {Nicklas}, \citenamefont {Zibold},
  \citenamefont {Bar-Gill}, \citenamefont {Kurizki},\ and\ \citenamefont
  {Oberthaler}}]{Gross2011}%
  \BibitemOpen
  \bibfield  {author} {\bibinfo {author} {\bibfnamefont {C.}~\bibnamefont
  {Gross}}, \bibinfo {author} {\bibfnamefont {H.}~\bibnamefont {Strobel}},
  \bibinfo {author} {\bibfnamefont {E.}~\bibnamefont {Nicklas}}, \bibinfo
  {author} {\bibfnamefont {T.}~\bibnamefont {Zibold}}, \bibinfo {author}
  {\bibfnamefont {N.}~\bibnamefont {Bar-Gill}}, \bibinfo {author}
  {\bibfnamefont {G.}~\bibnamefont {Kurizki}},\ and\ \bibinfo {author}
  {\bibfnamefont {M.~K.}\ \bibnamefont {Oberthaler}},\ }\bibfield  {title}
  {\bibinfo {title} {Atomic homodyne detection of continuous-variable entangled
  twin-atom states},\ }\href {https://doi.org/10.1038/nature10654} {\bibfield
  {journal} {\bibinfo  {journal} {Nature}\ }\textbf {\bibinfo {volume} {480}},\
  \bibinfo {pages} {219} (\bibinfo {year} {2011})}\BibitemShut {NoStop}%
\bibitem [{\citenamefont {Hamley}\ \emph {et~al.}(2012)\citenamefont {Hamley},
  \citenamefont {Gerving}, \citenamefont {Hoang}, \citenamefont {Bookjans},\
  and\ \citenamefont {Chapman}}]{Hamley2012}%
  \BibitemOpen
  \bibfield  {author} {\bibinfo {author} {\bibfnamefont {C.~D.}\ \bibnamefont
  {Hamley}}, \bibinfo {author} {\bibfnamefont {C.~S.}\ \bibnamefont {Gerving}},
  \bibinfo {author} {\bibfnamefont {T.~M.}\ \bibnamefont {Hoang}}, \bibinfo
  {author} {\bibfnamefont {E.~M.}\ \bibnamefont {Bookjans}},\ and\ \bibinfo
  {author} {\bibfnamefont {M.~S.}\ \bibnamefont {Chapman}},\ }\bibfield
  {title} {\bibinfo {title} {Spin-nematic squeezed vacuum in a quantum gas},\
  }\href {https://doi.org/10.1038/nphys2245} {\bibfield  {journal} {\bibinfo
  {journal} {Nature Phys.}\ }\textbf {\bibinfo {volume} {8}},\ \bibinfo {pages}
  {305} (\bibinfo {year} {2012})}\BibitemShut {NoStop}%
\bibitem [{\citenamefont {Zhang}\ and\ \citenamefont {Duan}(2013)}]{Zhang2013}%
  \BibitemOpen
  \bibfield  {author} {\bibinfo {author} {\bibfnamefont {Z.}~\bibnamefont
  {Zhang}}\ and\ \bibinfo {author} {\bibfnamefont {L.-M.}\ \bibnamefont
  {Duan}},\ }\bibfield  {title} {\bibinfo {title} {Generation of massive
  entanglement through an adiabatic quantum phase transition in a spinor
  condensate},\ }\href {https://doi.org/10.1103/PhysRevLett.111.180401}
  {\bibfield  {journal} {\bibinfo  {journal} {Phys. Rev. Lett.}\ }\textbf
  {\bibinfo {volume} {111}},\ \bibinfo {pages} {180401} (\bibinfo {year}
  {2013})}\BibitemShut {NoStop}%
\bibitem [{\citenamefont {Luo}\ \emph {et~al.}(2017)\citenamefont {Luo},
  \citenamefont {Zou}, \citenamefont {Wu}, \citenamefont {Liu}, \citenamefont
  {Han}, \citenamefont {Tey},\ and\ \citenamefont {You}}]{Luo2017}%
  \BibitemOpen
  \bibfield  {author} {\bibinfo {author} {\bibfnamefont {X.-Y.}\ \bibnamefont
  {Luo}}, \bibinfo {author} {\bibfnamefont {Y.-Q.}\ \bibnamefont {Zou}},
  \bibinfo {author} {\bibfnamefont {L.-N.}\ \bibnamefont {Wu}}, \bibinfo
  {author} {\bibfnamefont {Q.}~\bibnamefont {Liu}}, \bibinfo {author}
  {\bibfnamefont {M.-F.}\ \bibnamefont {Han}}, \bibinfo {author} {\bibfnamefont
  {M.~K.}\ \bibnamefont {Tey}},\ and\ \bibinfo {author} {\bibfnamefont
  {L.}~\bibnamefont {You}},\ }\bibfield  {title} {\bibinfo {title}
  {Deterministic entanglement generation from driving through quantum phase
  transitions},\ }\href {https://doi.org/10.1126/science.aag1106} {\bibfield
  {journal} {\bibinfo  {journal} {Science}\ }\textbf {\bibinfo {volume}
  {355}},\ \bibinfo {pages} {620} (\bibinfo {year} {2017})}\BibitemShut
  {NoStop}%
\bibitem [{\citenamefont {Feldmann}\ \emph {et~al.}(2018)\citenamefont
  {Feldmann}, \citenamefont {Gessner}, \citenamefont {Gabbrielli},
  \citenamefont {Klempt}, \citenamefont {Santos}, \citenamefont {Pezz\`e},\
  and\ \citenamefont {Smerzi}}]{Feldmann2018}%
  \BibitemOpen
  \bibfield  {author} {\bibinfo {author} {\bibfnamefont {P.}~\bibnamefont
  {Feldmann}}, \bibinfo {author} {\bibfnamefont {M.}~\bibnamefont {Gessner}},
  \bibinfo {author} {\bibfnamefont {M.}~\bibnamefont {Gabbrielli}}, \bibinfo
  {author} {\bibfnamefont {C.}~\bibnamefont {Klempt}}, \bibinfo {author}
  {\bibfnamefont {L.}~\bibnamefont {Santos}}, \bibinfo {author} {\bibfnamefont
  {L.}~\bibnamefont {Pezz\`e}},\ and\ \bibinfo {author} {\bibfnamefont
  {A.}~\bibnamefont {Smerzi}},\ }\bibfield  {title} {\bibinfo {title}
  {Interferometric sensitivity and entanglement by scanning through quantum
  phase transitions in spinor {Bose-Einstein} condensates},\ }\href
  {https://doi.org/10.1103/PhysRevA.97.032339} {\bibfield  {journal} {\bibinfo
  {journal} {Phys. Rev. A}\ }\textbf {\bibinfo {volume} {97}},\ \bibinfo
  {pages} {032339} (\bibinfo {year} {2018})}\BibitemShut {NoStop}%
\bibitem [{Note1()}]{Note1}%
  \BibitemOpen
  \bibinfo {note} {See Supplemental Material at [URL will be inserted by the
  publisher] for details on the quasi-adiabatic state preparation, including
  references~\cite {Zhang2013, Luo2017, Feldmann2018, Chang2004, Leslie2009,
  Klempt2010, Linnemann2016a}.}\BibitemShut {Stop}%
\bibitem [{\citenamefont {Castin}\ and\ \citenamefont
  {Dum}(1996)}]{Castin1996}%
  \BibitemOpen
  \bibfield  {author} {\bibinfo {author} {\bibfnamefont {Y.}~\bibnamefont
  {Castin}}\ and\ \bibinfo {author} {\bibfnamefont {R.}~\bibnamefont {Dum}},\
  }\bibfield  {title} {\bibinfo {title} {{Bose-Einstein} condensates in time
  dependent traps},\ }\href {https://doi.org/10.1103/PhysRevLett.77.5315}
  {\bibfield  {journal} {\bibinfo  {journal} {Phys. Rev. Lett.}\ }\textbf
  {\bibinfo {volume} {77}},\ \bibinfo {pages} {5315} (\bibinfo {year}
  {1996})}\BibitemShut {NoStop}%
\bibitem [{\citenamefont {Ammann}\ and\ \citenamefont
  {Christensen}(1997)}]{Ammann1997}%
  \BibitemOpen
  \bibfield  {author} {\bibinfo {author} {\bibfnamefont {H.}~\bibnamefont
  {Ammann}}\ and\ \bibinfo {author} {\bibfnamefont {N.}~\bibnamefont
  {Christensen}},\ }\bibfield  {title} {\bibinfo {title} {Delta kick cooling: A
  new method for cooling atoms},\ }\href
  {https://doi.org/10.1103/PhysRevLett.78.2088} {\bibfield  {journal} {\bibinfo
   {journal} {Phys. Rev. Lett.}\ }\textbf {\bibinfo {volume} {78}},\ \bibinfo
  {pages} {2088} (\bibinfo {year} {1997})}\BibitemShut {NoStop}%
\bibitem [{\citenamefont {Malia}\ \emph {et~al.}(2020)\citenamefont {Malia},
  \citenamefont {Mart\'{\i}nez-Rinc\'on}, \citenamefont {Wu}, \citenamefont
  {Hosten},\ and\ \citenamefont {Kasevich}}]{Malia2020}%
  \BibitemOpen
  \bibfield  {author} {\bibinfo {author} {\bibfnamefont {B.~K.}\ \bibnamefont
  {Malia}}, \bibinfo {author} {\bibfnamefont {J.}~\bibnamefont
  {Mart\'{\i}nez-Rinc\'on}}, \bibinfo {author} {\bibfnamefont {Y.}~\bibnamefont
  {Wu}}, \bibinfo {author} {\bibfnamefont {O.}~\bibnamefont {Hosten}},\ and\
  \bibinfo {author} {\bibfnamefont {M.~A.}\ \bibnamefont {Kasevich}},\
  }\bibfield  {title} {\bibinfo {title} {Free space ramsey spectroscopy in
  rubidium with noise below the quantum projection limit},\ }\href
  {https://doi.org/10.1103/PhysRevLett.125.043202} {\bibfield  {journal}
  {\bibinfo  {journal} {Phys. Rev. Lett.}\ }\textbf {\bibinfo {volume} {125}},\
  \bibinfo {pages} {043202} (\bibinfo {year} {2020})}\BibitemShut {NoStop}%
\bibitem [{\citenamefont {Berg}\ \emph {et~al.}(2015)\citenamefont {Berg},
  \citenamefont {Abend}, \citenamefont {Tackmann}, \citenamefont {Schubert},
  \citenamefont {Giese}, \citenamefont {Schleich}, \citenamefont {Narducci},
  \citenamefont {Ertmer},\ and\ \citenamefont {Rasel}}]{Berg2015}%
  \BibitemOpen
  \bibfield  {author} {\bibinfo {author} {\bibfnamefont {P.}~\bibnamefont
  {Berg}}, \bibinfo {author} {\bibfnamefont {S.}~\bibnamefont {Abend}},
  \bibinfo {author} {\bibfnamefont {G.}~\bibnamefont {Tackmann}}, \bibinfo
  {author} {\bibfnamefont {C.}~\bibnamefont {Schubert}}, \bibinfo {author}
  {\bibfnamefont {E.}~\bibnamefont {Giese}}, \bibinfo {author} {\bibfnamefont
  {W.}~\bibnamefont {Schleich}}, \bibinfo {author} {\bibfnamefont
  {F.}~\bibnamefont {Narducci}}, \bibinfo {author} {\bibfnamefont
  {W.}~\bibnamefont {Ertmer}},\ and\ \bibinfo {author} {\bibfnamefont
  {E.}~\bibnamefont {Rasel}},\ }\bibfield  {title} {\bibinfo {title}
  {Composite-light-pulse technique for high-precision atom interferometry},\
  }\href {https://doi.org/10.1103/physrevlett.114.063002} {\bibfield  {journal}
  {\bibinfo  {journal} {Phys. Rev. Lett.}\ }\textbf {\bibinfo {volume} {114}},\
  \bibinfo {pages} {063002} (\bibinfo {year} {2015})}\BibitemShut {NoStop}%
\bibitem [{\citenamefont {Butts}\ \emph {et~al.}(2013)\citenamefont {Butts},
  \citenamefont {Kotru}, \citenamefont {Kinast}, \citenamefont {Radojevic},
  \citenamefont {Timmons},\ and\ \citenamefont {Stoner}}]{Butts2013}%
  \BibitemOpen
  \bibfield  {author} {\bibinfo {author} {\bibfnamefont {D.~L.}\ \bibnamefont
  {Butts}}, \bibinfo {author} {\bibfnamefont {K.}~\bibnamefont {Kotru}},
  \bibinfo {author} {\bibfnamefont {J.~M.}\ \bibnamefont {Kinast}}, \bibinfo
  {author} {\bibfnamefont {A.~M.}\ \bibnamefont {Radojevic}}, \bibinfo {author}
  {\bibfnamefont {B.~P.}\ \bibnamefont {Timmons}},\ and\ \bibinfo {author}
  {\bibfnamefont {R.~E.}\ \bibnamefont {Stoner}},\ }\bibfield  {title}
  {\bibinfo {title} {Efficient broadband raman pulses for large-area atom
  interferometry},\ }\href {https://doi.org/10.1364/JOSAB.30.000922} {\bibfield
   {journal} {\bibinfo  {journal} {J. Opt. Soc. Am. B}\ }\textbf {\bibinfo
  {volume} {30}},\ \bibinfo {pages} {922} (\bibinfo {year} {2013})}\BibitemShut
  {NoStop}%
\bibitem [{\citenamefont {Kotru}\ \emph {et~al.}(2015)\citenamefont {Kotru},
  \citenamefont {Butts}, \citenamefont {Kinast},\ and\ \citenamefont
  {Stoner}}]{Kotru2015}%
  \BibitemOpen
  \bibfield  {author} {\bibinfo {author} {\bibfnamefont {K.}~\bibnamefont
  {Kotru}}, \bibinfo {author} {\bibfnamefont {D.~L.}\ \bibnamefont {Butts}},
  \bibinfo {author} {\bibfnamefont {J.~M.}\ \bibnamefont {Kinast}},\ and\
  \bibinfo {author} {\bibfnamefont {R.~E.}\ \bibnamefont {Stoner}},\ }\bibfield
   {title} {\bibinfo {title} {Large-area atom interferometry with
  frequency-swept raman adiabatic passage},\ }\href
  {https://doi.org/10.1103/PhysRevLett.115.103001} {\bibfield  {journal}
  {\bibinfo  {journal} {Phys. Rev. Lett.}\ }\textbf {\bibinfo {volume} {115}},\
  \bibinfo {pages} {103001} (\bibinfo {year} {2015})}\BibitemShut {NoStop}%
\bibitem [{\citenamefont {Jaffe}\ \emph {et~al.}(2018)\citenamefont {Jaffe},
  \citenamefont {Xu}, \citenamefont {Haslinger}, \citenamefont {M\"uller},\
  and\ \citenamefont {Hamilton}}]{Jaffe2018}%
  \BibitemOpen
  \bibfield  {author} {\bibinfo {author} {\bibfnamefont {M.}~\bibnamefont
  {Jaffe}}, \bibinfo {author} {\bibfnamefont {V.}~\bibnamefont {Xu}}, \bibinfo
  {author} {\bibfnamefont {P.}~\bibnamefont {Haslinger}}, \bibinfo {author}
  {\bibfnamefont {H.}~\bibnamefont {M\"uller}},\ and\ \bibinfo {author}
  {\bibfnamefont {P.}~\bibnamefont {Hamilton}},\ }\bibfield  {title} {\bibinfo
  {title} {Efficient adiabatic spin-dependent kicks in an atom
  interferometer},\ }\href {https://doi.org/10.1103/PhysRevLett.121.040402}
  {\bibfield  {journal} {\bibinfo  {journal} {Phys. Rev. Lett.}\ }\textbf
  {\bibinfo {volume} {121}},\ \bibinfo {pages} {040402} (\bibinfo {year}
  {2018})}\BibitemShut {NoStop}%
\bibitem [{Note2()}]{Note2}%
  \BibitemOpen
  \bibinfo {note} {Other techniques, such as Bragg transitions or Bloch
  oscillations~\cite {Kovachy2012, Abend2016}, can reach even better
  efficiencies but cannot be applied within our scheme.}\BibitemShut {Stop}%
\bibitem [{\citenamefont {Dimopoulos}\ \emph {et~al.}(2007)\citenamefont
  {Dimopoulos}, \citenamefont {Graham}, \citenamefont {Hogan},\ and\
  \citenamefont {Kasevich}}]{Dimopoulos2007}%
  \BibitemOpen
  \bibfield  {author} {\bibinfo {author} {\bibfnamefont {S.}~\bibnamefont
  {Dimopoulos}}, \bibinfo {author} {\bibfnamefont {P.~W.}\ \bibnamefont
  {Graham}}, \bibinfo {author} {\bibfnamefont {J.~M.}\ \bibnamefont {Hogan}},\
  and\ \bibinfo {author} {\bibfnamefont {M.~A.}\ \bibnamefont {Kasevich}},\
  }\bibfield  {title} {\bibinfo {title} {Testing general relativity with atom
  interferometry},\ }\href {https://doi.org/10.1103/PhysRevLett.98.111102}
  {\bibfield  {journal} {\bibinfo  {journal} {Phys. Rev. Lett.}\ }\textbf
  {\bibinfo {volume} {98}},\ \bibinfo {pages} {111102} (\bibinfo {year}
  {2007})}\BibitemShut {NoStop}%
\bibitem [{\citenamefont {Hartwig}\ \emph {et~al.}(2015)\citenamefont
  {Hartwig}, \citenamefont {Abend}, \citenamefont {Schubert}, \citenamefont
  {Schlippert}, \citenamefont {Ahlers}, \citenamefont {Posso-Trujillo},
  \citenamefont {Gaaloul}, \citenamefont {Ertmer},\ and\ \citenamefont
  {Rasel}}]{Hartwig2015}%
  \BibitemOpen
  \bibfield  {author} {\bibinfo {author} {\bibfnamefont {J.}~\bibnamefont
  {Hartwig}}, \bibinfo {author} {\bibfnamefont {S.}~\bibnamefont {Abend}},
  \bibinfo {author} {\bibfnamefont {C.}~\bibnamefont {Schubert}}, \bibinfo
  {author} {\bibfnamefont {D.}~\bibnamefont {Schlippert}}, \bibinfo {author}
  {\bibfnamefont {H.}~\bibnamefont {Ahlers}}, \bibinfo {author} {\bibfnamefont
  {K.}~\bibnamefont {Posso-Trujillo}}, \bibinfo {author} {\bibfnamefont
  {N.}~\bibnamefont {Gaaloul}}, \bibinfo {author} {\bibfnamefont
  {W.}~\bibnamefont {Ertmer}},\ and\ \bibinfo {author} {\bibfnamefont {E.~M.}\
  \bibnamefont {Rasel}},\ }\bibfield  {title} {\bibinfo {title} {Testing the
  universality of free fall with rubidium and ytterbium in a very large
  baseline atom interferometer},\ }\href
  {https://doi.org/10.1088/1367-2630/17/3/035011} {\bibfield  {journal}
  {\bibinfo  {journal} {New J. Phys.}\ }\textbf {\bibinfo {volume} {17}},\
  \bibinfo {pages} {035011} (\bibinfo {year} {2015})}\BibitemShut {NoStop}%
\bibitem [{\citenamefont {Asenbaum}\ \emph {et~al.}(2020)\citenamefont
  {Asenbaum}, \citenamefont {Overstreet}, \citenamefont {Kim}, \citenamefont
  {Curti},\ and\ \citenamefont {Kasevich}}]{Asenbaum2020}%
  \BibitemOpen
  \bibfield  {author} {\bibinfo {author} {\bibfnamefont {P.}~\bibnamefont
  {Asenbaum}}, \bibinfo {author} {\bibfnamefont {C.}~\bibnamefont
  {Overstreet}}, \bibinfo {author} {\bibfnamefont {M.}~\bibnamefont {Kim}},
  \bibinfo {author} {\bibfnamefont {J.}~\bibnamefont {Curti}},\ and\ \bibinfo
  {author} {\bibfnamefont {M.~A.}\ \bibnamefont {Kasevich}},\ }\bibfield
  {title} {\bibinfo {title} {Atom-interferometric test of the equivalence
  principle at the ${10}^{\ensuremath{-}12}$ level},\ }\href
  {https://doi.org/10.1103/PhysRevLett.125.191101} {\bibfield  {journal}
  {\bibinfo  {journal} {Phys. Rev. Lett.}\ }\textbf {\bibinfo {volume} {125}},\
  \bibinfo {pages} {191101} (\bibinfo {year} {2020})}\BibitemShut {NoStop}%
\bibitem [{\citenamefont {Schubert}\ \emph {et~al.}(2019)\citenamefont
  {Schubert}, \citenamefont {Schlippert}, \citenamefont {Abend}, \citenamefont
  {Giese}, \citenamefont {Roura}, \citenamefont {Schleich}, \citenamefont
  {Ertmer},\ and\ \citenamefont {Rasel}}]{Schubert2019}%
  \BibitemOpen
  \bibfield  {author} {\bibinfo {author} {\bibfnamefont {C.}~\bibnamefont
  {Schubert}}, \bibinfo {author} {\bibfnamefont {D.}~\bibnamefont
  {Schlippert}}, \bibinfo {author} {\bibfnamefont {S.}~\bibnamefont {Abend}},
  \bibinfo {author} {\bibfnamefont {E.}~\bibnamefont {Giese}}, \bibinfo
  {author} {\bibfnamefont {A.}~\bibnamefont {Roura}}, \bibinfo {author}
  {\bibfnamefont {W.~P.}\ \bibnamefont {Schleich}}, \bibinfo {author}
  {\bibfnamefont {W.}~\bibnamefont {Ertmer}},\ and\ \bibinfo {author}
  {\bibfnamefont {E.~M.}\ \bibnamefont {Rasel}},\ }\bibfield  {title} {\bibinfo
  {title} {Scalable, symmetric atom interferometer for infrasound gravitational
  wave detection},\ }\href@noop {} {\bibfield  {journal} {\bibinfo  {journal}
  {arXiv:1909.01951}\ } (\bibinfo {year} {2019})},\ \Eprint
  {https://arxiv.org/abs/1909.01951} {arXiv:1909.01951 [quant-ph]} \BibitemShut
  {NoStop}%
\bibitem [{\citenamefont {Dimopoulos}\ \emph {et~al.}(2008)\citenamefont
  {Dimopoulos}, \citenamefont {Graham}, \citenamefont {Hogan}, \citenamefont
  {Kasevich},\ and\ \citenamefont {Rajendran}}]{Dimopoulos2008}%
  \BibitemOpen
  \bibfield  {author} {\bibinfo {author} {\bibfnamefont {S.}~\bibnamefont
  {Dimopoulos}}, \bibinfo {author} {\bibfnamefont {P.~W.}\ \bibnamefont
  {Graham}}, \bibinfo {author} {\bibfnamefont {J.~M.}\ \bibnamefont {Hogan}},
  \bibinfo {author} {\bibfnamefont {M.~A.}\ \bibnamefont {Kasevich}},\ and\
  \bibinfo {author} {\bibfnamefont {S.}~\bibnamefont {Rajendran}},\ }\bibfield
  {title} {\bibinfo {title} {Atomic gravitational wave interferometric
  sensor},\ }\href {https://doi.org/10.1103/PhysRevD.78.122002} {\bibfield
  {journal} {\bibinfo  {journal} {Phys. Rev. D}\ }\textbf {\bibinfo {volume}
  {78}},\ \bibinfo {pages} {122002} (\bibinfo {year} {2008})}\BibitemShut
  {NoStop}%
\bibitem [{\citenamefont {Hogan}\ \emph {et~al.}(2011)\citenamefont {Hogan},
  \citenamefont {Johnson}, \citenamefont {Dickerson}, \citenamefont {Kovachy},
  \citenamefont {Sugarbaker}, \citenamefont {Chiow}, \citenamefont {Graham},
  \citenamefont {Kasevich}, \citenamefont {Saif}, \citenamefont {Rajendran},
  \citenamefont {Bouyer}, \citenamefont {Seery}, \citenamefont {Feinberg},\
  and\ \citenamefont {Keski-Kuha}}]{Hogan2011a}%
  \BibitemOpen
  \bibfield  {author} {\bibinfo {author} {\bibfnamefont {J.~M.}\ \bibnamefont
  {Hogan}}, \bibinfo {author} {\bibfnamefont {D.~M.~S.}\ \bibnamefont
  {Johnson}}, \bibinfo {author} {\bibfnamefont {S.}~\bibnamefont {Dickerson}},
  \bibinfo {author} {\bibfnamefont {T.}~\bibnamefont {Kovachy}}, \bibinfo
  {author} {\bibfnamefont {A.}~\bibnamefont {Sugarbaker}}, \bibinfo {author}
  {\bibfnamefont {S.-w.}\ \bibnamefont {Chiow}}, \bibinfo {author}
  {\bibfnamefont {P.~W.}\ \bibnamefont {Graham}}, \bibinfo {author}
  {\bibfnamefont {M.~A.}\ \bibnamefont {Kasevich}}, \bibinfo {author}
  {\bibfnamefont {B.}~\bibnamefont {Saif}}, \bibinfo {author} {\bibfnamefont
  {S.}~\bibnamefont {Rajendran}}, \bibinfo {author} {\bibfnamefont
  {P.}~\bibnamefont {Bouyer}}, \bibinfo {author} {\bibfnamefont {B.~D.}\
  \bibnamefont {Seery}}, \bibinfo {author} {\bibfnamefont {L.}~\bibnamefont
  {Feinberg}},\ and\ \bibinfo {author} {\bibfnamefont {R.}~\bibnamefont
  {Keski-Kuha}},\ }\bibfield  {title} {\bibinfo {title} {An atomic
  gravitational wave interferometric sensor in low earth orbit ({AGIS-LEO})},\
  }\href {http://rd.springer.com/article/10.1007/s10714-011-1182-x} {\bibfield
  {journal} {\bibinfo  {journal} {Gen. Relativ. Gravit.}\ }\textbf {\bibinfo
  {volume} {43}},\ \bibinfo {pages} {1953} (\bibinfo {year}
  {2011})}\BibitemShut {NoStop}%
\bibitem [{\citenamefont {Loriani}\ \emph {et~al.}(2019)\citenamefont
  {Loriani}, \citenamefont {Schlippert}, \citenamefont {Schubert},
  \citenamefont {Abend}, \citenamefont {Ahlers}, \citenamefont {Ertmer},
  \citenamefont {Rudolph}, \citenamefont {Hogan}, \citenamefont {Kasevich},
  \citenamefont {Rasel},\ and\ \citenamefont {Gaaloul}}]{Loriani2019}%
  \BibitemOpen
  \bibfield  {author} {\bibinfo {author} {\bibfnamefont {S.}~\bibnamefont
  {Loriani}}, \bibinfo {author} {\bibfnamefont {D.}~\bibnamefont {Schlippert}},
  \bibinfo {author} {\bibfnamefont {C.}~\bibnamefont {Schubert}}, \bibinfo
  {author} {\bibfnamefont {S.}~\bibnamefont {Abend}}, \bibinfo {author}
  {\bibfnamefont {H.}~\bibnamefont {Ahlers}}, \bibinfo {author} {\bibfnamefont
  {W.}~\bibnamefont {Ertmer}}, \bibinfo {author} {\bibfnamefont
  {J.}~\bibnamefont {Rudolph}}, \bibinfo {author} {\bibfnamefont {J.~M.}\
  \bibnamefont {Hogan}}, \bibinfo {author} {\bibfnamefont {M.~A.}\ \bibnamefont
  {Kasevich}}, \bibinfo {author} {\bibfnamefont {E.~M.}\ \bibnamefont
  {Rasel}},\ and\ \bibinfo {author} {\bibfnamefont {N.}~\bibnamefont
  {Gaaloul}},\ }\bibfield  {title} {\bibinfo {title} {Atomic source selection
  in space-borne gravitational wave detection},\ }\href
  {https://doi.org/10.1088/1367-2630/ab22d0} {\bibfield  {journal} {\bibinfo
  {journal} {New Journal of Physics}\ }\textbf {\bibinfo {volume} {21}},\
  \bibinfo {pages} {063030} (\bibinfo {year} {2019})}\BibitemShut {NoStop}%
\bibitem [{\citenamefont {El-Neaj}\ \emph {et~al.}(2020)\citenamefont
  {El-Neaj}, \citenamefont {Alpigiani}, \citenamefont {Amairi-Pyka},
  \citenamefont {Ara{\'u}jo}, \citenamefont {Bala{\v{z}}}, \citenamefont
  {Bassi}, \citenamefont {Bathe-Peters}, \citenamefont {Battelier},
  \citenamefont {Beli{\'{c}}}, \citenamefont {Bentine}, \citenamefont
  {Bernabeu}, \citenamefont {Bertoldi}, \citenamefont {Bingham}, \citenamefont
  {Blas}, \citenamefont {Bolpasi} \emph {et~al.}}]{ElNeaj2020}%
  \BibitemOpen
  \bibfield  {author} {\bibinfo {author} {\bibfnamefont {Y.~A.}\ \bibnamefont
  {El-Neaj}}, \bibinfo {author} {\bibfnamefont {C.}~\bibnamefont {Alpigiani}},
  \bibinfo {author} {\bibfnamefont {S.}~\bibnamefont {Amairi-Pyka}}, \bibinfo
  {author} {\bibfnamefont {H.}~\bibnamefont {Ara{\'u}jo}}, \bibinfo {author}
  {\bibfnamefont {A.}~\bibnamefont {Bala{\v{z}}}}, \bibinfo {author}
  {\bibfnamefont {A.}~\bibnamefont {Bassi}}, \bibinfo {author} {\bibfnamefont
  {L.}~\bibnamefont {Bathe-Peters}}, \bibinfo {author} {\bibfnamefont
  {B.}~\bibnamefont {Battelier}}, \bibinfo {author} {\bibfnamefont
  {A.}~\bibnamefont {Beli{\'{c}}}}, \bibinfo {author} {\bibfnamefont
  {E.}~\bibnamefont {Bentine}}, \bibinfo {author} {\bibfnamefont
  {J.}~\bibnamefont {Bernabeu}}, \bibinfo {author} {\bibfnamefont
  {A.}~\bibnamefont {Bertoldi}}, \bibinfo {author} {\bibfnamefont
  {R.}~\bibnamefont {Bingham}}, \bibinfo {author} {\bibfnamefont
  {D.}~\bibnamefont {Blas}}, \bibinfo {author} {\bibfnamefont {V.}~\bibnamefont
  {Bolpasi}}, \emph {et~al.},\ }\bibfield  {title} {\bibinfo {title} {Aedge:
  Atomic experiment for dark matter and gravity exploration in space},\ }\href
  {https://doi.org/10.1140/epjqt/s40507-020-0080-0} {\bibfield  {journal}
  {\bibinfo  {journal} {EPJ Quantum Technology}\ }\textbf {\bibinfo {volume}
  {7}},\ \bibinfo {pages} {6} (\bibinfo {year} {2020})}\BibitemShut {NoStop}%
\bibitem [{\citenamefont {Chang}\ \emph {et~al.}(2004)\citenamefont {Chang},
  \citenamefont {Hamley}, \citenamefont {Barrett}, \citenamefont {Sauer},
  \citenamefont {Fortier}, \citenamefont {Zhang}, \citenamefont {You},\ and\
  \citenamefont {Chapman}}]{Chang2004}%
  \BibitemOpen
  \bibfield  {author} {\bibinfo {author} {\bibfnamefont {M.-S.}\ \bibnamefont
  {Chang}}, \bibinfo {author} {\bibfnamefont {C.~D.}\ \bibnamefont {Hamley}},
  \bibinfo {author} {\bibfnamefont {M.~D.}\ \bibnamefont {Barrett}}, \bibinfo
  {author} {\bibfnamefont {J.~A.}\ \bibnamefont {Sauer}}, \bibinfo {author}
  {\bibfnamefont {K.~M.}\ \bibnamefont {Fortier}}, \bibinfo {author}
  {\bibfnamefont {W.}~\bibnamefont {Zhang}}, \bibinfo {author} {\bibfnamefont
  {L.}~\bibnamefont {You}},\ and\ \bibinfo {author} {\bibfnamefont {M.~S.}\
  \bibnamefont {Chapman}},\ }\bibfield  {title} {\bibinfo {title} {Observation
  of spinor dynamics in optically trapped $^{87}$rb bose-einstein
  condensates},\ }\href {https://doi.org/10.1103/PhysRevLett.92.140403}
  {\bibfield  {journal} {\bibinfo  {journal} {Phys. Rev. Lett.}\ }\textbf
  {\bibinfo {volume} {92}},\ \bibinfo {pages} {140403} (\bibinfo {year}
  {2004})}\BibitemShut {NoStop}%
\bibitem [{\citenamefont {Leslie}\ \emph {et~al.}(2009)\citenamefont {Leslie},
  \citenamefont {Guzman}, \citenamefont {Vengalattore}, \citenamefont {Sau},
  \citenamefont {Cohen},\ and\ \citenamefont {Stamper-Kurn}}]{Leslie2009}%
  \BibitemOpen
  \bibfield  {author} {\bibinfo {author} {\bibfnamefont {S.~R.}\ \bibnamefont
  {Leslie}}, \bibinfo {author} {\bibfnamefont {J.}~\bibnamefont {Guzman}},
  \bibinfo {author} {\bibfnamefont {M.}~\bibnamefont {Vengalattore}}, \bibinfo
  {author} {\bibfnamefont {J.~D.}\ \bibnamefont {Sau}}, \bibinfo {author}
  {\bibfnamefont {M.~L.}\ \bibnamefont {Cohen}},\ and\ \bibinfo {author}
  {\bibfnamefont {D.~M.}\ \bibnamefont {Stamper-Kurn}},\ }\bibfield  {title}
  {\bibinfo {title} {Amplification of fluctuations in a spinor {Bose-Einstein}
  condensate},\ }\href {https://doi.org/10.1103/PhysRevA.79.043631} {\bibfield
  {journal} {\bibinfo  {journal} {Phys. Rev. A}\ }\textbf {\bibinfo {volume}
  {79}},\ \bibinfo {eid} {043631} (\bibinfo {year} {2009})}\BibitemShut
  {NoStop}%
\bibitem [{\citenamefont {Klempt}\ \emph {et~al.}(2010)\citenamefont {Klempt},
  \citenamefont {Topic}, \citenamefont {Gebreyesus}, \citenamefont {Scherer},
  \citenamefont {Henninger}, \citenamefont {Hyllus}, \citenamefont {Ertmer},
  \citenamefont {Santos},\ and\ \citenamefont {Arlt}}]{Klempt2010}%
  \BibitemOpen
  \bibfield  {author} {\bibinfo {author} {\bibfnamefont {C.}~\bibnamefont
  {Klempt}}, \bibinfo {author} {\bibfnamefont {O.}~\bibnamefont {Topic}},
  \bibinfo {author} {\bibfnamefont {G.}~\bibnamefont {Gebreyesus}}, \bibinfo
  {author} {\bibfnamefont {M.}~\bibnamefont {Scherer}}, \bibinfo {author}
  {\bibfnamefont {T.}~\bibnamefont {Henninger}}, \bibinfo {author}
  {\bibfnamefont {P.}~\bibnamefont {Hyllus}}, \bibinfo {author} {\bibfnamefont
  {W.}~\bibnamefont {Ertmer}}, \bibinfo {author} {\bibfnamefont
  {L.}~\bibnamefont {Santos}},\ and\ \bibinfo {author} {\bibfnamefont {J.~J.}\
  \bibnamefont {Arlt}},\ }\bibfield  {title} {\bibinfo {title} {Parametric
  amplification of vacuum fluctuations in a spinor condensate},\ }\href
  {https://doi.org/10.1103/PhysRevLett.104.195303} {\bibfield  {journal}
  {\bibinfo  {journal} {Phys. Rev. Lett.}\ }\textbf {\bibinfo {volume} {104}},\
  \bibinfo {pages} {195303} (\bibinfo {year} {2010})}\BibitemShut {NoStop}%
\bibitem [{\citenamefont {Linnemann}\ \emph {et~al.}(2016)\citenamefont
  {Linnemann}, \citenamefont {Strobel}, \citenamefont {Muessel}, \citenamefont
  {Schulz}, \citenamefont {Lewis-Swan}, \citenamefont {Kheruntsyan},\ and\
  \citenamefont {Oberthaler}}]{Linnemann2016a}%
  \BibitemOpen
  \bibfield  {author} {\bibinfo {author} {\bibfnamefont {D.}~\bibnamefont
  {Linnemann}}, \bibinfo {author} {\bibfnamefont {H.}~\bibnamefont {Strobel}},
  \bibinfo {author} {\bibfnamefont {W.}~\bibnamefont {Muessel}}, \bibinfo
  {author} {\bibfnamefont {J.}~\bibnamefont {Schulz}}, \bibinfo {author}
  {\bibfnamefont {R.~J.}\ \bibnamefont {Lewis-Swan}}, \bibinfo {author}
  {\bibfnamefont {K.~V.}\ \bibnamefont {Kheruntsyan}},\ and\ \bibinfo {author}
  {\bibfnamefont {M.~K.}\ \bibnamefont {Oberthaler}},\ }\bibfield  {title}
  {\bibinfo {title} {Quantum-enhanced sensing based on time reversal of
  nonlinear dynamics},\ }\href {https://doi.org/10.1103/PhysRevLett.117.013001}
  {\bibfield  {journal} {\bibinfo  {journal} {Phys. Rev. Lett.}\ }\textbf
  {\bibinfo {volume} {117}},\ \bibinfo {pages} {013001} (\bibinfo {year}
  {2016})}\BibitemShut {NoStop}%
\bibitem [{\citenamefont {Kovachy}\ \emph {et~al.}(2012)\citenamefont
  {Kovachy}, \citenamefont {Chiow},\ and\ \citenamefont
  {Kasevich}}]{Kovachy2012}%
  \BibitemOpen
  \bibfield  {author} {\bibinfo {author} {\bibfnamefont {T.}~\bibnamefont
  {Kovachy}}, \bibinfo {author} {\bibfnamefont {S.-w.}\ \bibnamefont {Chiow}},\
  and\ \bibinfo {author} {\bibfnamefont {M.~A.}\ \bibnamefont {Kasevich}},\
  }\bibfield  {title} {\bibinfo {title} {Adiabatic-rapid-passage multiphoton
  bragg atom optics},\ }\href {https://doi.org/10.1103/PhysRevA.86.011606}
  {\bibfield  {journal} {\bibinfo  {journal} {Phys. Rev. A}\ }\textbf {\bibinfo
  {volume} {86}},\ \bibinfo {pages} {011606(R)} (\bibinfo {year}
  {2012})}\BibitemShut {NoStop}%
\bibitem [{\citenamefont {Abend}\ \emph {et~al.}(2016)\citenamefont {Abend},
  \citenamefont {Gebbe}, \citenamefont {Gersemann}, \citenamefont {Ahlers},
  \citenamefont {M{\"u}ntinga}, \citenamefont {Giese}, \citenamefont {Gaaloul},
  \citenamefont {Schubert}, \citenamefont {L{\"a}mmerzahl}, \citenamefont
  {Ertmer}, \citenamefont {Schleich},\ and\ \citenamefont {Rasel}}]{Abend2016}%
  \BibitemOpen
  \bibfield  {author} {\bibinfo {author} {\bibfnamefont {S.}~\bibnamefont
  {Abend}}, \bibinfo {author} {\bibfnamefont {M.}~\bibnamefont {Gebbe}},
  \bibinfo {author} {\bibfnamefont {M.}~\bibnamefont {Gersemann}}, \bibinfo
  {author} {\bibfnamefont {H.}~\bibnamefont {Ahlers}}, \bibinfo {author}
  {\bibfnamefont {H.}~\bibnamefont {M{\"u}ntinga}}, \bibinfo {author}
  {\bibfnamefont {E.}~\bibnamefont {Giese}}, \bibinfo {author} {\bibfnamefont
  {N.}~\bibnamefont {Gaaloul}}, \bibinfo {author} {\bibfnamefont
  {C.}~\bibnamefont {Schubert}}, \bibinfo {author} {\bibfnamefont
  {C.}~\bibnamefont {L{\"a}mmerzahl}}, \bibinfo {author} {\bibfnamefont
  {W.}~\bibnamefont {Ertmer}}, \bibinfo {author} {\bibfnamefont {W.~P.}\
  \bibnamefont {Schleich}},\ and\ \bibinfo {author} {\bibfnamefont {E.~M.}\
  \bibnamefont {Rasel}},\ }\bibfield  {title} {\bibinfo {title} {Atom-chip
  fountain gravimeter},\ }\href
  {https://doi.org/10.1103/physrevlett.117.203003} {\bibfield  {journal}
  {\bibinfo  {journal} {Phys. Rev. Lett.}\ }\textbf {\bibinfo {volume} {117}},\
  \bibinfo {pages} {203003} (\bibinfo {year} {2016})}\BibitemShut {NoStop}%
\end{thebibliography}%

\pagebreak
\clearpage
\widetext
\begin{center}
\textbf{\large Momentum entanglement for atom interferometry (Supplemental Material)}
\end{center}
\setcounter{equation}{0}
\setcounter{figure}{0}
\setcounter{table}{0}
\makeatletter
\renewcommand{\theequation}{S\arabic{equation}}
\renewcommand{\thefigure}{S\arabic{figure}}

\section{Quasi-adiabatic state preparation}
\begin{figure*}[b!]
\centering
\includegraphics[]{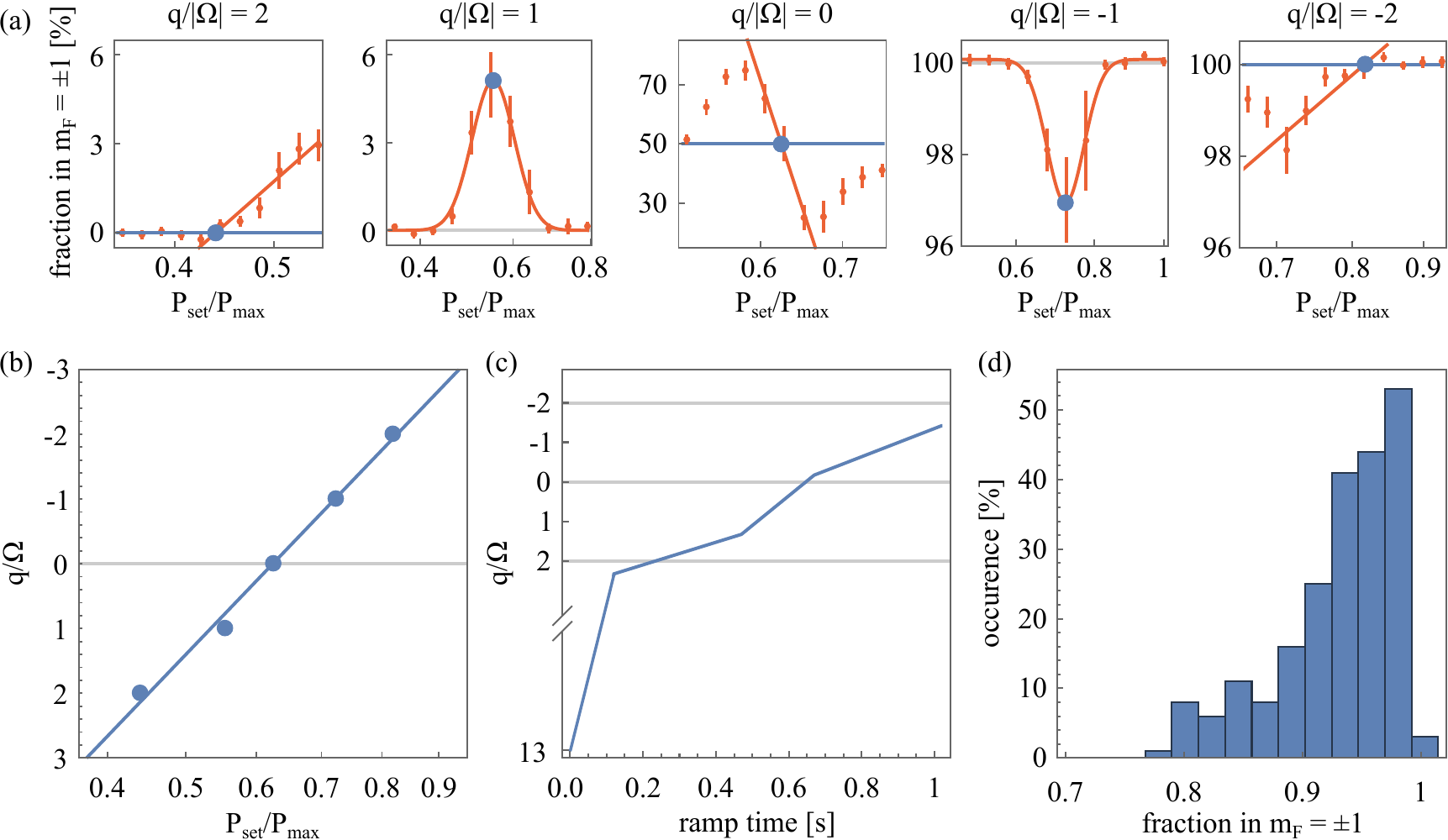}
\caption{
Calibration of $q$ and details of the quasi-adiabatic ramp.
(a) We conduct five independent measurements to determine the position of characteristic $q/|\Omega|$ values in the relevant parameter regime.
(b) We obtain $q/|\Omega|$ as a function of the MW dressing power.
(c) Temporal shape of the experimentally optimized ramp. In the quasi-adiabatic case, the crossing of the second quantum phase transition is not necessary to reach largely populated twin-Fock states.
(d) Histogram of the fraction of atoms transferred into the levels $\ket{1,\pm 1}$.
}
\label{figRamp}
\end{figure*}
Here, we present details on our entangled state preparation.
Compared to earlier work~\cite{Zhang2013, Luo2017, Feldmann2018}, we extended the method to calibrate the relevant parameter $q/|\Omega|$ and realized a slightly faster procedure with comparable performance.
We employ spin-changing collisions to generate entanglement in a spin-$1$ Bose-Einstein condensates \cite{Chang2004, Leslie2009, Klempt2010, Linnemann2016a}.
Within the single-mode approximation and the magnetization free subspace, $\ev{N_1 - N_{-1}} =0$, the dynamics is governed by
\begin{align}\label{eq1}
H = \frac{\Omega}{2N} \left( 2(  a^\dagger_{0} a^\dagger_{0} a_{1} a_{-1} + a^\dagger_{1} a^\dagger_{-1} a_{0} a_{0}) 
+ (2\hat{N}_0 -1)(\hat{N}_1 + \hat{N}_{-1}) \right) 
+ q(\hat{N}_1 + \hat{N}_{-1}).
\end{align}
Here, $N=N_{-1}+N_0+N_1$ is the total (preserved) number of atoms in the three levels $\ket{1,-1/0/+1}$, $\Omega$ represents the interaction strength and $q$ is the effective energy that an atom in the levels $\ket{1,\pm 1}$ has with respect to an atom in $\ket{1,0}$.
The first term represents spin-changing collisions which produce correlated pairs in the levels $\ket{1,\pm 1}$. The second term expresses spin-preserving collisions, and the third term describes the effective energy of the levels due to the quadratic Zeeman shift and the microwave (MW) dressing.

The many-body ground-state of the system is determined by the value of $q/|\Omega|$, which is initially at $q/|\Omega| = 13$.
The interaction strength  $|\Omega| = h\times$\SI{3}{\Hz} is measured independently.
A pair of atoms in the levels $\ket{1,\pm 1}$ has thus a higher energy than a pair in the level $\ket{1,0}$.
Therefore, the initially prepared state with all atoms in $\ket{1,0}$ constitutes the ground state.
At $q/|\Omega| =\pm2$, the system undergoes two quantum phase transitions (QPT).
For $q/|\Omega| < -2$ the system is in the twin-Fock phase, where the state with a symmetric population of the levels $\ket{1,\pm 1}$ and an empty level $\ket{1,0}$ represents the ground state.
Ramping $q$ adiabatically from the polar phase $q/|\Omega| > 2$ to the twin-Fock phase $q/|\Omega| < -2$, the system perfectly follows the ground state, populates the levels $\ket{1,\pm 1}$ by spin-changing collisions and finally produces a clean twin-Fock state.

We change the value of $q$ by varying the intensity of the MW-dressing field.
In the case without dressing, our magnetic field results in  $q=h\times$\SI{38.5}{\Hz} due to the quadratic Zeeman shift only.
The quasi-adiabatic state preparation requires a calibration of $q/|\Omega|$ as a function of the MW power.
Figure~\ref{figRamp} (a) shows the single measurements from which we obtain the calibration, which follow the same scheme:
We prepare an initial spin configuration and apply the MW dressing field for a certain duration.
During this time, the spin-changing dynamics is enabled, thus transferring atoms between the levels $\ket{1,0}$ and $\ket{1,\pm 1}$.
After the dynamics, the population of the spin levels is measured.
For a determination of $q/|\Omega|=2$, the BEC is prepared in the level $\ket{1,0}$ and the MW dressing is applied for \SI{90}{ms}.
The relative MW power of $P_\text{set}=0.45 P_\text{max}$ marks the threshold, where spin dynamics starts, and corresponds to the quantum phase transition (QPT) from the polar phase to the phase of broken axisymmetry.
The point $q/|\Omega|=1$ is marked by a maximal transfer after a duration of \SI{110}{ms}.
Note that in this case, the third term in Eq.~(\ref{eq1}) cancels the second one, because $\Omega<0$ and initially $N=N_0$.
To determine the MW power that corresponds to $q/|\Omega|=0$, the condensate is prepared with $50\%$ of the atoms in $\ket{1,0}$ and $25\%$ in $\ket{1,\pm 1}$, respectively, via symmetric radiofrequency coupling.
For $q/|\Omega|\gtrsim 0$, the atoms tend to be predominantly transferred to $\ket{1,\pm 1}$ and reverse for $q/|\Omega|\lesssim 0$.
Directly at $q/|\Omega|=0$, the population remains equally distributed.
The chosen evolution time for this measurement is \SI{60}{ms}.
To achieve a good estimation of $q/|\Omega|$ as a function of the MW power, $q/|\Omega|=-2$ and $q/|\Omega|=-1$ are investigated equivalently to their positive counterparts, but with the initial condensate prepared symmetrically in $\ket{1,\pm 1}$.

While an adiabatic state preparation is optimal for slow passages of the QPTs, losses and heating require a compromise with a quasi-adiabatic procedure.
A transfer fidelity near 1 is not required in our case, as we remove the residual atoms in $\ket{1,0}$ anyhow.
We employ a combination of four linear ramps in $q$ (Figure~\ref{figRamp}~(c)).
Within $120$~ms, we quickly ramp to $q/ |\Omega|= 2.4$. The QPT at $q/ |\Omega|= 2$ is slowly passed within $350$~ms.
After the crossing the ramping speed is slightly increased.
For linear ramps, the population of the levels $\ket{1,\pm 1}$ oscillates during the ramping and actually reaches a maximum before the second QPT~\cite{Luo2017}.
To save ramping time, we therefore stop the ramp at $q/|\Omega|=-1.6$ before the second QPT.
The presented parameters were experimentally optimized to reach an efficient mean transfer above \SI{90}{\percent} in the shortest possible time.
The distribution of the final fraction of atoms transferred into the twin-Fock state is shown in figure~\ref{figRamp}~(d) and yields a transfer of \SI{93(5)}{\percent}.

\end{document}